\documentclass[times,sort&compress,3p]{elsarticle}
\usepackage[labelfont=bf]{caption}

\usepackage{amsmath,amsfonts,amssymb,amsthm,booktabs,color,epsfig,graphicx,hyperref,url}

\usepackage{multicol}
\usepackage{multirow}
\usepackage{algorithm}
\usepackage{algorithmic}

\theoremstyle{plain}% Theorem-like structures provided by amsthm.sty
\newtheorem{theorem}{Theorem}

\theoremstyle{definition}

\begin{document}

\begin{frontmatter}

\title{Type I multivariate Pólya-Aeppli distributions with applications}

\author[1]{Claire Geldenhuys \corref{mycorrespondingauthor}}
\author[1]{Rene Ehlers}
\author[1]{Andriette Bekker}
\address[1]{Department of Statistics, University of Pretoria, Pretoria, South Africa}

\cortext[mycorrespondingauthor]{Corresponding author. Email address: \url{clairecgeldenhuys@gmail.com}}

\begin{abstract}
An extensive body of literature exists that specifically addresses the univariate case of zero-inflated count models. In contrast, research pertaining to multivariate models is notably less developed. We proposed two new parsimonious multivariate models which can be used to model correlated multivariate overdispersed count data. Furthermore, for different parameter settings and sample sizes, various simulations are performed. In conclusion, we demonstrated the performance of the newly proposed multivariate candidates on two benchmark datasets, which surpasses that of several alternative approaches.\\

\end{abstract}


\end{frontmatter}

\section{Introduction\label{sec:1}}

Multivariate count data is frequently observed in the fields of medicine, sports, insurance, genomics and other domains \cite{ghosh2021new}, \cite{liu2019new} and \cite{tian2018type}. Although several distributions exist to model univariate and bivariate count data, the multivariate case is a field that is still in much need of development. The presence of overdispersion, zero-inflation and zero-and-one inflation in count data occurs regularly in practice and multivariate models must be flexible enough to take this into account.

The univariate Pólya-Aeppli distribution \cite{Anscombe1950}, which is elaborated upon in \cite{Johnson2005} as a compound Poisson
distribution, finds extensive applicability in risk theory, particularly in the estimation of ruin probability and numerous other domains  \cite{minkova2004polya}. An important characteristic of the Pólya-Aeppli distribution is its overdispersion, which distinguishes it from the equi-dispersed Poisson distribution often used for modeling count data. The Pólya-Aeppli distribution therefore has increased flexibility compared to the Poisson distribution, as data in practice is often overdispersed.

Interested in extending the properties of the univariate Pólya-Aeppli
distribution to the bivariate case, \cite{Minkova2014a} proposed
the Type I bivariate Pólya-Aeppli (Type I BPA) distribution. The latter is a compound bivariate Poisson distribution with a geometric compounding distribution, which is overdispersed with respect to the bivariate Poisson distibution.

One of the challenges of the Type I BPA distribution proposed in \cite{Minkova2014a}, is that the probability mass function (PMF) was obtained through the use of recursive formulas. This poses challenges in expanding the distribution to the multivariate case. Additionally, it complicates maximum likelihood estimation (MLE) for the parameters due to the fact that the likelihood function is also given in terms of recursive formulas. This results in likelihood estimation also being computationally intensive \cite{balakrishnan2017likelihood}.

In this paper we extend the Type I BPA distribution to the Type I multivariate Pólya-Aeppli (Type I MPA) distribution and propose an alternative approach to obtaining the PMF through the use of Laguerre polynomials. Expressing the PMF of the univariate Pólya-Aeppli distribution in terms of Laguerre polynomials, was first introduced by \cite{Galliher1959}.

In order to create distributions with even more flexibility with respect to the Poisson distribution and therefore also more flexibility with respect to the univariate Pólya-Aeppli distribution, \cite{minkova2013} proposed weighted versions of the univariate Pólya-Aeppli distribution. One such model is the Pólya-Aeppli distribution with non-negative weight function $w(z)=\frac{1}{z+1}$, $z\in\{0,1,\ldots\}$ applied to the Poisson random variable. This particular distribution is overdispersed with respect to the univariate Pólya-Aeppli distribution. We are interested
in extending the properties of this weighted univariate Pólya-Aeppli distribution to the multivariate
case as this will allow us to more accurately model data that has
significant overdispersion, and that the Type I MPA
distribution cannot account for. We therefore introduce the Type I weighted multivariate Pólya-Aeppli (Type I WMPA)
distribution with non-negative weight functions $w(z_{i})=\frac{1}{z_{i}+1}$, $z_{i}\in\{0,1,\ldots\}$ applied to the Poisson random variables, therefore a compound weighted multivariate Poisson distribution with a geometric compounding distribution. The  Type I WMPA distribution is then overdispersed with respect to the
Type I MPA distribution.

We will also consider the application of both the Type I MPA and Type I WMPA distributions to overdispered zero-inflated and zero-and-one inflated count data. While other distributions have been suggested for modeling this kind of data, our proposed distributions provide a substantial contribution to the existing literature. In all examples we examine, they demonstrate a superior fit compared to other models that have been evaluated.

This work adds to the existing literature in several dimensions.  Firstly, we present two new parsimonious multivariate count models. Secondly, we obtain the expressions for the PMFs in terms of Laguerre polynomials and not via recursive formulas, this enjoys a simpler structure and method for computation of the MLE process. Thirdly, these new candidates are worthy contenders in the playfield of overdispersed multivariate count data.

The rest of this paper is structured as follows; in Section \hyperref[sec:2]{2} and \hyperref[sec:3]{3} we introduce the Type I MPA and Type I WMPA distributions respectively. We provide their properties, PMFs as well as method of moments (MoM) estimates. In Section \hyperref[sec:4]{4} we  discuss the process of MLE for these distributions and in Section \hyperref[sec:5]{5} we consider their dispersion through use of a generalized dispersion index (GDI) \cite{kokonendji2018fisher}. In Section \hyperref[sec:6]{6} we conduct a simulation study in order to evaluate the performance of the different estimation methods and in Section \hyperref[sec:7]{7} we apply these distributions to zero-inflated and zero-and-one inflated count data. We present a conclusion and provide insightful observations regarding our proposed distributions in Section \hyperref[sec:8]{8}.

\section{Type I multivariate Pólya-Aeppli distribution}\label{sec:2}

To extend the univariate Pólya-Aeppli distribution to the Type I BPA distribution, \cite{Minkova2014a} proposed compounding a bivariate Poisson distribution with the geometric distribution via the trivariate reduction method. We will follow a similar approach in order to extend the Type I BPA distribution to the Type I MPA distribution.

Suppose we have independent Poisson distributed random variables $Z_{i}\sim Poi\left(\lambda_{i}\right)$,
$i\in\{1,\ldots,k\}$ and $Z_{k+1}\sim Poi\left(\lambda_{k+1}\right)$.
Also suppose $U_{ij}$ and $W_{i\ell}$ , $\ensuremath{i\in\{1,\ldots,k\}}$,
$\ensuremath{j\in\{1,\ldots,Z_{i}\}}$ and $\ensuremath{\ell\in\{1,\ldots,Z_{k+1}\}}$
are independent identically distributed $Ge_{1}\left(1-\rho\right)$
random variables, all independent of the Poisson random variables
$Z_{i}$, $i\in\{1,\ldots,k\}$ and $Z_{k+1}$. Let $\begin{alignedat}{1}U_{i}=\sum_{j=1}^{Z_{i}}U_{ij}\quad & W_{i}=\sum_{\ell=1}^{Z_{k+1}}W_{i\ell}\end{alignedat}
$. Define $N_{i}$, $\ensuremath{i\in\{1,\ldots,k\}}$
such that $\begin{alignedat}{1}N_{i} & =U_{i}+W_{i}\end{alignedat}
$. The joint distribution of $\left(N_{1},\ldots,N_{k}\right)$ is
then said to have a Type I MPA distribution
denoted by $\left(N_{1},\ldots,N_{k}\right)\sim MPA_{I}\left(\lambda_{1},\ldots,\lambda_{k},\lambda_{k+1},\rho\right)$. Since $U_{i}\sim PA\left(\lambda_{i},\rho\right)$ the probability
generating function (PGF) of $U_{i}$, $\ensuremath{i\in\{1,\ldots,k\}}$, \cite{Johnson2005} is

{\large{}
\begin{flalign}
\begin{alignedat}{1}\text{E}\left(s_{i}^{U_{i}}\right)\end{alignedat}
 & =e^{-\lambda_{i}\left(1-\psi_{1}\left(s_{i}\right)\right)}\label{eq:1}
\end{flalign}
}{\large\par}

where $\begin{alignedat}{1}\psi_{1}\left(s_{i}\right) & =\frac{\left(1-\rho\right)s_{i}}{1-\rho s_{i}}.\end{alignedat}
$ The expression for the PMF of the Pólya-Aeppli distribution as a function of Laguerre polynomials, was first given in \cite{Galliher1959}

{\large{}
\begin{equation}
\begin{alignedat}{1}\text{Pr}\left(U_{i}=n_{i}\right) & \begin{cases}
\begin{array}{cc}
e^{-\lambda_{i}}, & n_{i}=0\\
\frac{e^{-\lambda_{i}}\lambda_{i}\left(1-\rho\right)\rho^{n_{i}-1}}{n_{i}}L_{n_{i}-1}^{1}\left(x_{i}\right), & n_{i}=1,2,\ldots,
\end{array}\end{cases}\end{alignedat}
\label{eq:2}
\end{equation}
}{\large\par}

where $x_{i}=\frac{-\lambda_{i}\left(1-\rho\right)}{\rho}$, $i\in\{1,\ldots,k\}$
and $L_{n_{i}}^{\alpha}\left(x_{i}\right)$ is the Laguerre polynomial in (\ref{A27}). In this paper, the Laguerre polynomials will be employed to represent the PMFs of the distributions. Furthermore,
since the $W_{i\ell}$s are independent of each other and independent
of $Z_{k+1}$ it follows that

{\large{}
\[
\begin{alignedat}{1}\text{E}\left(\prod_{i=1}^{k}s_{i}^{W_{i}}\right) & =\text{E}\left(\prod_{i=1}^{k}s_{i}^{\sum_{\ell=1}^{z_{k+1}}W_{i\ell}}\right)\\
 & =\text{E}_{Z_{k+1}}\left[\left.\text{E}\left(\prod_{i=1}^{k}\prod_{\ell=1}^{z_{k+1}}s_{i}^{W_{i\ell}}\right)\right|Z_{k+1}\right]\\
 & =\text{E}_{Z_{k+1}}\left[\prod_{i=1}^{k}\prod_{\ell=1}^{z_{k+1}}\text{E}\left(s_{i}^{W_{i\ell}}\right)\right]\\
 & =\text{E}_{Z_{k+1}}\left[\left(\prod_{i=1}^{k}\psi_{1}\left(s_{i}\right)\right)^{Z_{k+1}}\right]\\
 & =e^{-\lambda_{k+1}\left(1-\prod_{i=1}^{k}\psi_{1}\left(s_{i}\right)\right)}.
\end{alignedat}
\]
}{\large\par}

The joint PGF of $\left(N_{1},\ldots,N_{k}\right)$ is then given
by {\large{}
\begin{flalign}
\begin{alignedat}{1}\psi_{N_{1},\ldots,N_{k}}\left(s_{1},\ldots,s_{k}\right) & =\left(\prod_{i=1}^{k}\text{E}\left(s_{i}^{U_{i}}\right)\right)\left(\text{E}\left(\prod_{i=1}^{k}s_{i}^{W_{i}}\right)\right)\\
 & =e^{-\lambda_{k+1}\left(1-\prod_{i=1}^{k}\psi_{1}\left(s_{i}\right)\right)}\prod_{i=1}^{k}e^{-\lambda_{i}\left(1-\psi_{1}\left(s_{i}\right)\right)}.
\end{alignedat}
\label{eq:3}
\end{flalign}
}{\large\par}
This can be observed as an expansion of the bivariate PGF in \cite{Minkova2014a}.
To obtain the moments of $\left(N_{1},\ldots,N_{k}\right)\sim MPA_{I}\left(\lambda_{1},\ldots,\lambda_{k},\lambda_{k+1},\rho\right)$,
we can calculate the first, second and joint partial derivatives of
the PGF in (\ref{eq:3}) with respect to $\left\{ s_{1},\ldots,s_{k}\right\} $.
For $\ensuremath{i\in\{1,\ldots,k\}}$, $\ensuremath{j\in\{1,\ldots,k\}}$,
$i\neq j$, it follows that

{\large{}
\begin{flalign}
\begin{alignedat}{1}\text{E}\left(N_{i}\right) & =\frac{\lambda_{i}+\lambda_{k+1}}{\left(1-\rho\right)}\\
\text{E}\left(N_{i}N_{j}\right) & =\frac{\left(\lambda_{i}+\lambda_{k+1}\right)\left(\lambda_{j}+\lambda_{k+1}\right)+\lambda_{k+1}}{\left(1-\rho\right)^{2}}\\
\text{E}\left(N_{i}^{2}\right) & =\frac{2\rho\left(\lambda_{i}+\lambda_{k+1}\right)}{\left(1-\rho\right)^{2}}+\left(\frac{\lambda_{i}+\lambda_{k+1}}{1-\rho}\right)^{2}\\
\text{Var}\left(N_{i}\right) & =\frac{\left(1+\rho\right)\left(\lambda_{i}+\lambda_{k+1}\right)}{\left(1-\rho\right)^{2}}\\
\text{Cov}\left(N_{i},N_{j}\right) & =\frac{\lambda_{k+1}}{\left(1-\rho\right)^{2}}\\
\text{Corr}\left(N_{i},N_{j}\right)= & \frac{\lambda_{k+1}}{\left(1+\rho\right)\sqrt{\left(\lambda_{i}+\lambda_{k+1}\right)\left(\lambda_{j}+\lambda_{k+1}\right)}}.
\end{alignedat}
\label{eq:4}
\end{flalign}
}{\large\par}

\subsection{Probability mass function}
\begin{theorem}
The PMF of $\left(N_{1},\ldots,N_{k}\right)\sim MPA_{I}\left(\lambda_{1},\ldots,\lambda_{k},\lambda_{k+1},\rho\right)$
is

{\large{}
\begin{flalign*}
\begin{alignedat}{1}f\left(0,\ldots,0\right) & =e^{-\lambda_{k+1}}\prod_{i=1}^{k}e^{-\lambda_{i}}\\
f\left(n_{a},\ldots,n_{b},0,\ldots,0\right) & =\left(\prod_{i=a}^{b}\frac{\lambda_{i}\rho^{n_{i}-1}\left(1-\rho\right)}{n_{i}}L_{n_{i}-1}^{1}\left(x_{i}\right)\right)f\left(0,\ldots,0\right),\\
 & \quad\left\{ n_{1},\ldots,n_{k}\right\} \equiv\left\{ n_{a},\ldots,n_{b},n_{c},\ldots,n_{d}\right\} ,\\
 & \quad\left\{ n_{a},\ldots,n_{b}\right\} \neq\left\{ 0,\ldots,0\right\} ,a\leq b,\\
 & \quad\left\{ n_{c},\ldots,n_{d}\right\} =\left\{ 0,\ldots,0\right\} ,c\leq d\\
f\left(n_{1},\ldots,n_{k}\right) & =\left[\left(\prod_{i=1}^{k}\frac{\lambda_{i}\rho^{n_{i}-1}\left(1-\rho\right)}{n_{i}}L_{n_{i}-1}^{1}\left(x_{i}\right)\right)\right.\\
 & \left.\quad+\sum_{r=1}^{min(n_{1},...,n_{k})}\frac{\lambda_{k+1}^{r}}{r!}\left(\prod_{i=1}^{k}(1-\rho)^{r}\rho^{n_{i}-r}L_{n_{i}-r}^{r-1}\left(x_{i}\right)\right)\right]f\left(0,\ldots,0\right),\\
 & \quad\left\{ n_{1},\ldots,n_{k}\right\} \neq\left\{ 0,\ldots,0\right\}, 
\end{alignedat}
\end{flalign*}
}{\large\par}

where $x_{i}=\frac{-\lambda_{i}\left(1-\rho\right)}{\rho}$, $i\in\{1,\ldots,k\}$
and $L_{n_{i}}^{\alpha}\left(x_{i}\right)$ is the Laguerre polynomial
in (\ref{A27}).
\end{theorem}

\begin{proof}[\textbf{\upshape Proof:}]
For the first part of the proof, $\left\{ n_{1},\ldots,n_{k}\right\} =\left\{ 0,\ldots,0\right\} $.
It follows from \cite{Kocherlakota1992} and (\ref{eq:3}) that

{\large{}
\[
\begin{alignedat}{1}f\left(0,\ldots,0\right) & =\psi_{N_{1},\ldots,N_{k}}\left(0,\ldots,0\right)=e^{-\lambda_{k+1}}\prod_{i=1}^{k}e^{-\lambda_{i}}.\end{alignedat}
\]
}{\large\par}

The second part of the proof is derived for the case where at least
one, but not all of the $\left\{ n_{1},\ldots,n_{k}\right\} $ is
zero, with the rest being greater than zero. Consider $\left\{ n_{1},\ldots,n_{k}\right\} $
such that $\left\{ n_{1},\ldots,n_{k}\right\} \equiv\left\{ n_{a},\ldots,n_{b},n_{c},\ldots,n_{d}\right\} $,
where $\left\{ n_{a},\ldots,n_{b}\right\} \neq\left\{ 0,\ldots,0\right\} $,
$a\leq b$ and $\left\{ n_{c},\ldots,n_{d}\right\} =\left\{ 0,\ldots,0\right\} $,
$c\leq d$. Since $U_{i}\sim PA\left(\lambda_{i},\rho\right)$, $i\in\{1,\ldots,k\}$,
from (\ref{eq:2}), (\ref{eq:3}) and (\ref{A1}) it follows that

{\large{}
\begin{flalign*}
\begin{alignedat}{1}\psi_{N_{1},\ldots,N_{k}}(s_{a},\ldots,s_{b},0,\ldots,0) & =\sum_{n_{a},\ldots,n_{b}=1}^{\infty}f\left(n_{a},\ldots,n_{b},0,\ldots,0\right)s_{a}^{n_{a}}\cdots s_{b}^{n_{b}}\\
 & =e^{-\lambda_{k+1}}\left(\prod_{i=a}^{b}e^{-\lambda_{i}\left(1-\psi_{1}\left(s_{i}\right)\right)}\right)\left(\prod_{j=c}^{d}e^{-\lambda_{j}}\right)\\
 & =e^{-\lambda_{k+1}}\left(\prod_{i=a}^{b}\psi_{U_{i}}(s_{i})\right)\left(\prod_{j=c}^{d}e^{-\lambda_{j}}\right)\\
 & =\sum_{n_{a},\ldots,n_{b}=1}^{\infty}\left[e^{-\lambda_{k+1}}\left(\prod_{i=a}^{b}\text{Pr}\left(U_{i}=n_{i}\right)\right)\left(\prod_{j=c}^{d}e^{-\lambda_{j}}\right)\right]s_{a}^{n_{a}}\cdots s_{b}^{n_{b}}\\
 & =\sum_{n_{a},\ldots,n_{b}=1}^{\infty}\left[\prod_{i=a}^{b}\frac{\lambda_{i}\rho^{n_{i}-1}\left(1-\rho\right)}{n_{i}}L_{n_{i}-1}^{1}\left(x_{i}\right)f\left(0,\ldots,0\right)\right]s_{a}^{n_{a}}\cdots s_{b}^{n_{b}},
\end{alignedat}
\end{flalign*}
}{\large\par}

where $x_{i}=\frac{-\lambda_{i}\left(1-\rho\right)}{\rho}$, $i\in\{a,\ldots,b\}$.
The result for the second part of the proof follows from this.

The third and final part of the proof the PMF is derived when all
of the $\left\{ n_{1},\ldots,n_{k}\right\} $ is greater than zero.
Let $\left\{ n_{1},\ldots,n_{k}\right\} \neq\left\{ 0,\ldots,0\right\} $.
Let $x_{i}=\frac{-\lambda_{i}\left(1-\rho\right)}{\rho}$ and
$z_{i}=\rho s_{i}$ , $i\in\{1,\ldots,k\}$ then by using (\ref{A1}), we
can write the PGF\textit{ }in (\ref{eq:3}) as

{\large{}
\begin{flalign*}
\begin{alignedat}{1}\psi_{N_{1},\ldots,N_{k}}(s_{1},\ldots,s_{k}) & =\sum_{n_{1},\ldots,n_{k}=1}^{\infty}f\left(n_{1},\ldots,n_{k}\right)s_{1}^{n_{1}}\cdots s_{k}^{n_{k}}\\
 & =\left[e^{\lambda_{k+1}\prod_{i=1}^{k}\frac{(1-\rho)s_{i}}{1-\rho s_{i}}}\prod_{i=1}^{k}e^{\lambda_{i}\left(\frac{(1-\rho)s_{i}}{1-\rho s_{i}}\right)}\right]f\left(0,\ldots,0\right)\\
 & =\left[e^{\lambda_{k+1}\prod_{i=1}^{k}\frac{(1-\rho)}{\rho}\frac{z_{i}}{1-z_{i}}}\prod_{i=1}^{k}e^{\left(\frac{x_{i}z_{i}}{z_{i}-1}\right)}\right]f\left(0,\ldots,0\right)\\
 & =\left[\left(\prod_{i=1}^{k}e^{\left(\frac{x_{i}z_{i}}{z_{i}-1}\right)}\right)+\sum_{r=1}^{\infty}\frac{\lambda_{k+1}^{r}}{r!}\left(\prod_{i=1}^{k}\frac{(1-\rho)^{r}}{\rho^{r}}\frac{z_{i}^{r}}{\left(1-z_{i}\right)^{r}}e^{\left(\frac{x_{i}z_{i}}{z_{i}-1}\right)}\right)\right]f\left(0,\ldots,0\right)\\
 & =\psi_{N_{1},\ldots,N_{k}}^{\left(1\right)}\left(s_{1},\ldots,s_{k}\right)+\psi_{N_{1},\ldots,N_{k}}^{\left(2\right)}\left(s_{1},\ldots,s_{k}\right).
\end{alignedat}
\end{flalign*}
}{\large\par}

The expression for $\psi_{N_{1},\ldots,N_{k}}^{\left(1\right)}(s_{1},\ldots,s_{k})$
can be rewritten using (\ref{eq:1}) and (\ref{eq:2}), such that

{\large{}
\begin{flalign*}
\begin{alignedat}{1}\psi_{N_{1},\ldots,N_{k}}^{\left(1\right)}(s_{1},\ldots,s_{k}) & =\sum_{n_{1},\ldots,n_{k}=1}^{\infty}f^{\left(1\right)}\left(n_{1},\ldots,n_{k}\right)s_{1}^{n_{1}}\cdots s_{k}^{n_{k}}\\
 & =\left[\prod_{i=1}^{k}e^{\left(\frac{x_{i}z_{i}}{z_{i}-1}\right)}\right]f\left(0,\ldots,0\right)\\
 & =e^{-\lambda_{k+1}}\prod_{i=1}^{k}e^{-\lambda_{i}\left(1-\psi_{1}\left(s_{i}\right)\right)}\\
 & =e^{-\lambda_{k+1}}\prod_{i=1}^{k}\psi_{U_{i}}(s_{i})\\
 & =\sum_{n_{1},\ldots,n_{k}=1}^{\infty}\left[e^{-\lambda_{k+1}}\prod_{i=1}^{k}\text{Pr}\left(U_{i}=n_{i}\right)\right]s_{1}^{n_{1}}\cdots s_{k}^{n_{k}}.\\
 & =\sum_{n_{1},\ldots,n_{k}=1}^{\infty}\left[\left(\prod_{i=1}^{k}\frac{\lambda_{i}\left(1-\rho\right)\rho^{n_{i}-1}}{n_{i}}L_{n_{i}-1}^{1}\left(x_{i}\right)\right)f\left(0,\ldots,0\right)\right]s_{1}^{n_{1}}\cdots s_{k}^{n_{k}}.
\end{alignedat}
\end{flalign*}
}{\large\par}

Using the generating function for Laguerre polynomials in (\ref{A26}),
as well as the mathematical results in (\ref{A1}), $\psi_{N_{1},\ldots,N_{k}}^{\left(2\right)}(s_{1},\ldots,s_{k})$
can be rewritten as

{\large{}
\begin{flalign*}
\begin{alignedat}{1}\psi_{N_{1},\ldots,N_{k}}^{\left(2\right)}(s_{1},\ldots,s_{k}) & =\sum_{n_{1},\ldots,n_{k}=1}^{\infty}f^{\left(2\right)}\left(n_{1},\ldots,n_{k}\right)s_{1}^{n_{1}}\cdots s_{k}^{n_{k}}\\
 & =\left[\sum_{r=1}^{\infty}\frac{\lambda_{k+1}^{r}}{r!}\left(\prod_{i=1}^{k}\frac{(1-\rho)^{r}}{\rho^{r}}\frac{z_{i}^{r}}{\left(1-z_{i}\right)^{r}}e^{\left(\frac{x_{i}z_{i}}{z_{i}-1}\right)}\right)\right]f\left(0,\ldots,0\right)\\
 & =\left[\sum_{r=0}^{\infty}\frac{\lambda_{k+1}^{r+1}}{\left(r+1\right)!}\prod_{i=1}^{k}\frac{(1-\rho)^{r+1}}{\rho^{r+1}}z_{i}^{r+1}\sum_{n_{i}=0}^{\infty}L_{n_{i}}^{r}\left(x_{i}\right)z_{i}^{n_{i}}\right]f\left(0,\ldots,0\right)\\
 & =\left[\sum_{n_{1},\ldots,n_{k}=1}^{\infty}\sum_{r=1}^{min(n_{1},...,n_{k})}\frac{\lambda_{k+1}^{r}}{r!}\prod_{i=1}^{k}\frac{(1-\rho)^{r}}{\rho^{r}}L_{n_{i}-r}^{r-1}\left(x_{i}\right)z_{i}^{n_{i}}\right]f\left(0,\ldots,0\right)\\
 & =\sum_{n_{1},\ldots,n_{k}=1}^{\infty}\left[\left(\sum_{r=1}^{min(n_{1},...,n_{k})}\frac{\lambda_{k+1}^{r}}{r!}\prod_{i=1}^{k}(1-\rho)^{r}\rho^{n_{i}-r}L_{n_{i}-r}^{r-1}\left(x_{i}\right)\right)f\left(0,\ldots,0\right)\right]s_{1}^{n_{1}}\cdots s_{k}^{n_{k}}.
\end{alignedat}
\end{flalign*}
}{\large\par}

The PMF can then be calculated as

{\large{}
\begin{flalign*}
\begin{alignedat}{1}f\left(n_{1},\ldots,n_{k}\right) & =f^{\left(1\right)}\left(n_{1},\ldots,n_{k}\right)+f^{\left(2\right)}\left(n_{1},\ldots,n_{k}\right).\end{alignedat}
\end{flalign*}
}{\large\par}

\end{proof}

\subsection{Method of moments estimates}

Let $\left(N_{1},\ldots,N_{k}\right)\sim MPA_{I}\left(\lambda_{1},\ldots,\lambda_{k},\lambda_{k+1},\rho\right)$
and consider a random sample of size $m$ from this distribution.
The observed values can then be given as $\left(n_{1\ell},\ldots,n_{k\ell}\right)$
where $\ell\in\{1,\ldots,m\}$. In addition, for $i,j\in\{1,\ldots,k\}$,
$i\neq j$, let

{\large{}
\begin{flalign*}
\begin{alignedat}{1}\overline{n}_{i} & =\frac{1}{m}\sum_{\ell=1}^{m}n_{i\ell}\\
s_{ij} & =\frac{1}{\left(m-1\right)}\sum_{\ell=1}^{m}\left(n_{i\ell}-\overline{n}_{i}\right)\left(n_{j\ell}-\overline{n}_{j}\right)\\
s_{i}^{2} & =\frac{1}{\left(m-1\right)}\sum_{\ell=1}^{m}\left(n_{i\ell}-\overline{n}_{i}\right)^{2}.
\end{alignedat}
\end{flalign*}
}{\large\par}

Using these together with the expressions in (\ref{eq:4}) for $\text{E}\left(N_{i}\right)$,
$\text{Cov}\left(N_{i},N_{j}\right)$ and $\text{Var}\left(N_{i}\right)$,
the MoM estimates are given as

{\large{}
\begin{flalign*}
\begin{alignedat}{1}\hat{\lambda}_{i} & =\left(1-\hat{\rho}\right)\overline{n}_{i}-\left(1-\hat{\rho}\right)^{2}\frac{\left(k-2\right)!}{k!}\sum_{i,j=1,i\neq j}^{k}s_{ij}\\
\hat{\lambda}_{k+1} & =\left(1-\hat{\rho}\right)^{2}\frac{\left(k-2\right)!}{k!}\sum_{i,j=1,i\neq j}^{k}s_{ij}\\
\hat{\rho} & =\frac{\sum_{i=1}^{k}s_{i}^{2}-\sum_{i=1}^{k}\overline{n}_{i}}{\sum_{i=1}^{k}s_{i}^{2}+\sum_{i=1}^{k}\overline{n}_{i}},
\end{alignedat}
\end{flalign*}
}{\large\par}

where $i,j\in\{1,\ldots,k\}$, $i\neq j$. This can be easily observed as an expansion of the moment estimates for the Type I BPA distribution in \cite{Minkova2014a}.

\section{Type I weighted multivariate Pólya-Aeppli distribution}\label{sec:3}

In \cite{minkova2013} a weighted version of the univariate
Pólya-Aeppli distribution was introduced with non-negative weight
function $w(z)=\frac{1}{z+1}$, $z\in\{0,1,\ldots\}$. This distribution
is overdispersed with respect to the Pólya-Aeppli distribution. We will extend it in a similar way as we extended the Type I MPA to propose a new multivariate distribution, the  Type I WMPA that is overdispersed with respect to the Type I MPA distribution.

Suppose we have independent distributed random variables $Z_{i}^{w}\sim Poi\left(\lambda_{i}\right)$,
$\ensuremath{i\in\{1,\ldots,k\}}$ and $Z_{k+1}^{w}\sim Poi\left(\lambda_{k+1}\right)$.
Let $Z_{i}^{w}$ and $Z_{k+1}^{w}$ be the weighted version of $Z_{i}$
and $Z_{k+1}$ respectively, with non-negative weight functions $w(z_{i})=\frac{1}{z_{i}+1}$,
$z_{i}\in\{0,1,\ldots\}$. Also suppose $U_{ij}^{w}$ and $W_{i\ell}^{w}$
, $\ensuremath{i\in\{1,\ldots,k\}}$, $\ensuremath{j\in\{1,\ldots,Z_{i}^{w}\}}$
and $\ensuremath{\ell\in\{1,\ldots,Z_{k+1}^{w}\}}$ are independent
and identically distributed $Ge_{1}\left(1-\rho\right)$ random variables,
all independent of the Poisson random variables $Z_{i}^{w}$, $i\in\{1,\ldots,k\}$
and $Z_{k+1}^{w}$. Let $\begin{alignedat}{1}U_{i}^{w}=\sum_{j=1}^{Z_{i}^{w}}U_{ij}^{w}\quad & W_{i}^{w}=\sum_{\ell=1}^{Z_{k+1}^{w}}W_{i\ell}^{w}\end{alignedat}
$. Define $N_{i}^{w}$, $\ensuremath{i\in\{1,\ldots,k\}}$
such that $\begin{alignedat}{1}N_{i}^{w} & =U_{i}^{w}+W_{i}^{w}\end{alignedat}
$. The joint distribution of $\left(N_{1}^{w},\ldots,N_{k}^{w}\right)$
is then said to have a Type I WMPA distribution
denoted by $\left(N_{1}^{w},\ldots,N_{k}^{w}\right)\sim WMPA_{I}\left(\lambda_{1},\ldots,\lambda_{k},\lambda_{k+1},\rho\right)$.
Since $U_{i}^{w}\sim WPA\left(\lambda_{i},\rho\right)$, with non-negative
weight functions $w(z_{i})=\frac{1}{z_{i}+1}$, $z_{i}\in\{0,1,\ldots\}$,
the PGF of $U_{i}^{w}$, $\ensuremath{i\in\{1,\ldots,k\}}$, \cite{minkova2013} is

{\large{}
\begin{flalign}
\begin{alignedat}{1}\text{E}\left(s_{i}^{U_{i}^{w}}\right) & =\frac{e^{-\lambda_{i}}}{1-e^{-\lambda_{i}}}\frac{\left(e^{\lambda_{i}\psi_{1}\left(s_{i}\right)}-1\right)}{\psi_{1}\left(s_{i}\right)},\end{alignedat}
\label{eq:5}
\end{flalign}
}{\large\par}

where $\begin{alignedat}{1}\psi_{1}\left(s_{i}\right) & =\frac{\left(1-\rho\right)s_{i}}{1-\rho s_{i}}.\end{alignedat}
$ The expression for the PMF for $U_{i}^{w}\sim WPA\left(\lambda_{i},\rho\right)$
with non-negative weight functions $w(z_{i})=\frac{1}{z_{i}+1}$,
$z_{i}\in\{0,1,\ldots\}$, \cite{minkova2013} can be given as a function of the Laguerre polynomials in (\ref{A27}) and (\ref{eq:A31}) such that

{\large{}
\begin{equation}
\begin{alignedat}{1}\text{Pr}\left(U_{i}^{w}=n_{i}\right) & \begin{cases}
\begin{array}{cc}
\frac{\lambda_{i}e^{-\lambda_{i}}}{\left(1-e^{-\lambda_{i}}\right)}, & n_{i}=0\\
\frac{e^{-\lambda_{i}}}{\left(1-e^{-\lambda_{i}}\right)}\frac{\lambda_{i}^{2}\left(1-\rho\right)\rho^{n_{i}-1}}{n_{i}\left(n_{i}+1\right)}L_{n_{i}-1}^{2}\left(x_{i}\right), & n_{i}=1,2,\ldots,
\end{array}\end{cases}\end{alignedat}
\label{eq:6}
\end{equation}
}{\large\par}

where $x_{i}=\frac{-\lambda_{i}\left(1-\rho\right)}{\rho}$, $\ensuremath{i\in\{1,\ldots,k\}}$
and $L_{n_{i}}^{\alpha}\left(x_{i}\right)$ is the Laguerre polynomial in (\ref{A27}). Furthermore, since the $W_{i\ell}^{w}$s
are independent of each other and independent of $Z_{k+1}^{w}$ it
follows similarly as in (\ref{eq:3}) that {\large{}
\[
\begin{alignedat}{1}\text{E}\left(\prod_{i=1}^{k}s_{i}^{W_{i}^{w}}\right) & =\frac{e^{-\lambda_{k+1}}}{\left(1-e^{-\lambda_{k+1}}\right)}\frac{\left(e^{\lambda_{k+1}\prod_{i=1}^{k}\psi_{1}\left(s_{i}\right)}-1\right)}{\prod_{i=1}^{k}\psi_{1}\left(s_{i}\right)}.\end{alignedat}
\]
}

The joint PGF of $\left(N_{1}^{w},\ldots,N_{k}^{w}\right)$ follows as {\large{}
\begin{flalign}
\begin{alignedat}{1}\psi_{N_{1}^{w},...,N_{k}^{w}}(s_{1},...,s_{k}) & =\left[\prod_{i=1}^{k}\text{E}\left(s_{i}^{U_{i}^{w}}\right)\right]\left[\text{E}\left(\prod_{i=1}^{k}s_{i}^{W_{i}^{w}}\right)\right]\\
 & =\frac{e^{-\lambda_{k+1}}}{\left(1-e^{-\lambda_{k+1}}\right)}\left(e^{\lambda_{k+1}\prod_{i=1}^{k}\psi_{1}\left(s_{i}\right)}-1\right)\prod_{i=1}^{k}\frac{e^{-\lambda_{i}}}{\left(1-e^{-\lambda_{i}}\right)}\frac{\left(e^{\lambda_{i}\psi_{1}\left(s_{i}\right)}-1\right)}{\psi_{1}\left(s_{i}\right)^{2}}.
\end{alignedat}
\label{eq:7}
\end{flalign}
}{\large\par}

The moments of $\left(N_{1}^{w},\ldots,N_{k}^{w}\right)\sim WMVPA_{I}\left(\lambda_{1},\ldots,\lambda_{k},\lambda_{k+1},\rho\right)$
can be obtained similarly as in (\ref{eq:4}). For $\ensuremath{i\in\{1,\ldots,k\}}$,
$\ensuremath{j\in\{1,\ldots,k\}}$, $i\neq j$ it follows that

{\large{}
\begin{flalign}
\begin{alignedat}{1}\text{E}\left(N_{i}^{w}\right) & =\frac{\lambda_{i}}{(1-\rho)\left(1-e^{-\lambda_{i}}\right)}+\frac{\lambda_{k+1}}{(1-\rho)\left(1-e^{-\lambda_{k+1}}\right)}-\frac{2}{(1-\rho)}\\
\text{E}\left(N_{i}^{w}N_{j}^{w}\right) & =\left[\frac{\lambda_{i}}{(1-\rho)\left(1-e^{-\lambda_{i}}\right)}+\frac{\lambda_{k+1}}{(1-\rho)\left(1-e^{-\lambda_{k+1}}\right)}-\frac{2}{(1-\rho)}\right]\\
 & \times\left[\frac{\lambda_{j}}{(1-\rho)\left(1-e^{-\lambda_{j}}\right)}+\frac{\lambda_{k+1}}{(1-\rho)\left(1-e^{-\lambda_{k+1}}\right)}-\frac{2}{(1-\rho)}\right]\\
 & +\frac{\lambda_{k+1}}{(1-\rho)^{2}\left(1-e^{-\lambda_{k+1}}\right)}\left[1-\frac{\lambda_{k+1}e^{-\lambda_{k+1}}}{\left(1-e^{-\lambda_{k+1}}\right)}\right]\\
\text{E}\left(N_{i}^{w^{2}}\right) & =\frac{2\rho\lambda_{i}}{(1-\rho)^{2}\left(1-e^{-\lambda_{i}}\right)}+\frac{2\rho\lambda_{k+1}}{(1-\rho)^{2}\left(1-e^{-\lambda_{k+1}}\right)}\\
 & -\frac{\lambda_{i}^{2}e^{-\lambda_{i}}}{(1-\rho)^{2}\left(1-e^{-\lambda_{i}}\right)^{2}}-\frac{\lambda_{k+1}^{2}e^{-\lambda_{k+1}}}{(1-\rho)^{2}\left(1-e^{-\lambda_{k+1}}\right)^{2}}+\frac{2\left(1-2\rho\right)}{(1-\rho)^{2}}\\
 & +\left[\frac{\lambda_{i}}{(1-\rho)\left(1-e^{-\lambda_{i}}\right)}+\frac{\lambda_{k+1}}{(1-\rho)\left(1-e^{-\lambda_{k+1}}\right)}-\frac{2}{(1-\rho)}\right]^{2}\\
\text{Var}\left(N_{i}^{w}\right) & =\frac{\left(1+\rho\right)\lambda_{i}}{(1-\rho)^{2}\left(1-e^{-\lambda_{i}}\right)}+\frac{\left(1+\rho\right)\lambda_{k+1}}{(1-\rho)^{2}\left(1-e^{-\lambda_{k+1}}\right)}\\
 & -\frac{\lambda_{i}^{2}e^{-\lambda_{i}}}{(1-\rho)^{2}\left(1-e^{-\lambda_{i}}\right)^{2}}-\frac{\lambda_{k+1}^{2}e^{-\lambda_{k+1}}}{(1-\rho)^{2}\left(1-e^{-\lambda_{k+1}}\right)^{2}}-\frac{2\rho}{(1-\rho)^{2}}\\
\text{Cov}\left(N_{i}^{w},N_{j}^{w}\right) & =\frac{\lambda_{k+1}}{(1-\rho)^{2}\left(1-e^{-\lambda_{k+1}}\right)}\left[1-\frac{\lambda_{k+1}e^{-\lambda_{k+1}}}{\left(1-e^{-\lambda_{k+1}}\right)}\right]\\
\text{Corr}(N_{i}^{w},N_{j}^{w}) & =\frac{\text{Cov}\left(N_{i}^{w},N_{j}^{w}\right)}{\sqrt{\text{Var}\left(N_{i}^{w}\right)}\sqrt{\text{Var}\left(N_{j}^{w}\right)}}.
\end{alignedat}
\label{eq:8}
\end{flalign}
}{\large\par}

\subsection{Probability Mass Function}

\begin{theorem} 
$ $
The PMF of $\left(N_{1}^{w},\ldots,N_{k}^{w}\right)\sim WMPA_{I}\left(\lambda_{1},\ldots,\lambda_{k},\lambda_{k+1},\rho\right)$
is

{\large{}
\begin{flalign*}
\begin{alignedat}{1}f^{w}\left(0,\ldots,0\right) & =\frac{\lambda_{k+1}e^{-\lambda_{k+1}}}{\left(1-e^{-\lambda_{k+1}}\right)}\prod_{i=1}^{k}\frac{\lambda_{i}e^{-\lambda_{i}}}{\left(1-e^{-\lambda_{i}}\right)}\\
f^{w}\left(n_{a},\ldots,n_{b},0,\ldots,0\right) & =\left(\prod_{i=a}^{b}\frac{\lambda_{i}(1-\rho)\rho^{n_{i}-1}}{n_{i}\left(n_{i}+1\right)}L_{n_{i}-1}^{2}\left(x_{i}\right)\right)f^{w}\left(0,\ldots,0\right),\\
 & \quad\left\{ n_{1},\ldots,n_{k}\right\} \equiv\left\{ n_{a},\ldots,n_{b},n_{c},\ldots,n_{d}\right\} ,\\
 & \quad\left\{ n_{a},\ldots,n_{b}\right\} \neq\left\{ 0,\ldots,0\right\} ,a\leq b,\\
 & \quad\left\{ n_{c},\ldots,n_{d}\right\} =\left\{ 0,\ldots,0\right\} ,c\leq d\\
f^{w}\left(n_{1},...,n_{k}\right) & =\left[\left(\prod_{i=1}^{k}\frac{\lambda_{i}(1-\rho)\rho^{n_{i}-1}}{n_{i}\left(n_{i}+1\right)}L_{n_{i}-1}^{2}\left(x_{i}\right)\right)+\frac{\lambda_{k+1}}{2!}\left(\prod_{i=1}^{k}\frac{(1-\rho)\rho^{n_{i}-1}}{n_{i}}L_{n_{i}-1}^{1}\left(x_{i}\right)\right)\right.\\
 & \left.\quad+\sum_{r=1}^{min(n_{1},...,n_{k})}\frac{\lambda_{k+1}^{r+1}}{\left(r+2\right)!}\left(\prod_{i=1}^{k}\frac{(1-\rho)^{r}\rho^{n_{i}-r}}{\lambda_{i}}\left(L_{n_{i}-r}^{r-1}\left(x_{i}\right)-\left(\begin{array}{c}
n_{i}-1\\
n_{i}-r
\end{array}\right)\right)\right)\right]\\
 & \quad\times f^{w}\left(0,\ldots,0\right),\\
 & \quad\left\{ n_{1},\ldots,n_{k}\right\} \neq\left\{ 0,\ldots,0\right\} 
\end{alignedat}
\end{flalign*}
}{\large\par}

where $x_{i}=\frac{-\lambda_{i}\left(1-\rho\right)}{\rho}$, $i\in\{1,\ldots,k\}$
and $L_{n_{i}}^{\alpha}\left(x_{i}\right)$ is the Laguerre polynomial
in (\ref{A27}).
$ $
\end{theorem}

\begin{proof}[\textbf{\upshape Proof:}]
$ $
For the first part of the proof, $\left\{ n_{1},\ldots,n_{k}\right\} =\left\{ 0,\ldots,0\right\} $.
From \cite{Kocherlakota1992} and (\ref{A1}), we can rewrite the PGF in (\ref{eq:7})
as

{\large{}
\begin{flalign*}
\begin{alignedat}{1}f^{w}\left(0,\ldots,0\right) & =\psi_{N_{1}^{w},\ldots,N_{k}^{w}}(0,\ldots,0)\\
 & =\frac{e^{-\lambda_{k+1}}}{\left(1-e^{-\lambda_{k+1}}\right)}\left[\lambda_{k+1}+\sum_{m_{k+1}=2}^{\infty}\frac{\lambda_{k+1}^{m_{k+1}}\prod_{i=1}^{k}\psi_{1}\left(s_{i}\right)^{m_{k+1}-1}}{m_{k+1}!}\right]\\
 & \times\prod_{i=1}^{k}\frac{e^{-\lambda_{i}}}{\left(1-e^{-\lambda_{i}}\right)}\left[\lambda_{i}+\sum_{m_{i}=2}^{\infty}\frac{\lambda_{i}^{m_{i}}\psi_{1}\left(s_{i}\right)^{m_{i}-1}}{m_{i}!}\right]\\
 & =\frac{\lambda_{k+1}e^{-\lambda_{k+1}}}{\left(1-e^{-\lambda_{k+1}}\right)}\prod_{i=1}^{k}\frac{\lambda_{i}e^{-\lambda_{i}}}{\left(1-e^{-\lambda_{i}}\right)}.
\end{alignedat}
\end{flalign*}
}{\large\par}

The second part of the proof is derived for the case where at least
one, but not all of the $\left\{ n_{1},\ldots,n_{k}\right\} $ is
zero, with the rest being greater than zero. Consider $\left\{ n_{1},\ldots,n_{k}\right\} $
such that $\left\{ n_{1},\ldots,n_{k}\right\} \equiv\left\{ n_{a},\ldots,n_{b},n_{c},\ldots,n_{d}\right\} $,
where $\left\{ n_{a},\ldots,n_{b}\right\} \neq\left\{ 0,\ldots,0\right\} $,
$a\leq b$ and $\left\{ n_{c},\ldots,n_{d}\right\} =\left\{ 0,\ldots,0\right\} $,
$c\leq d$. Since $U_{i}^{w}\sim WPA\left(\lambda_{i},\rho\right)$,
$i\in\{1,\ldots,k\}$, from (\ref{eq:5}), (\ref{eq:6}) and (\ref{eq:7}),
it follows that

{\large{}
\begin{flalign*}
\begin{alignedat}{1} & \psi_{N_{1}^{w},\ldots,N_{k}^{w}}(s_{a},\ldots,s_{b},0,\ldots,0)\\
 & =\sum_{n_{a},\ldots,n_{b}=1}^{\infty}f^{w}\left(n_{a},\ldots,n_{b},0,\ldots,0\right)s_{a}^{n_{a}}\cdots s_{b}^{n_{b}}\\
 & =\frac{\lambda_{k+1}e^{-\lambda_{k+1}}}{\left(1-e^{-\lambda_{k+1}}\right)}\left(\prod_{i=a}^{b}\frac{e^{-\lambda_{i}}}{\left(1-e^{-\lambda_{i}}\right)}\frac{\left(e^{\lambda_{i}\psi_{1}\left(s_{i}\right)}-1\right)}{\psi_{1}\left(s_{i}\right)}\right)\left(\prod_{j=c}^{d}\frac{\lambda_{j}e^{-\lambda_{j}}}{\left(1-e^{-\lambda_{j}}\right)}\right)\\
 & =\frac{\lambda_{k+1}e^{-\lambda_{k+1}}}{\left(1-e^{-\lambda_{k+1}}\right)}\left(\prod_{i=a}^{b}\psi_{U_{i}^{w}}(s_{i})\right)\left(\prod_{j=c}^{d}\frac{\lambda_{j}e^{-\lambda_{j}}}{\left(1-e^{-\lambda_{j}}\right)}\right)\\
 & =\sum_{n_{a},\ldots,n_{b}=1}^{\infty}\left[\frac{\lambda_{k+1}e^{-\lambda_{k+1}}}{\left(1-e^{-\lambda_{k+1}}\right)}\left(\prod_{i=a}^{b}\text{Pr}\left(U_{i}^{w}=n_{i}\right)\right)\left(\prod_{j=c}^{d}\frac{\lambda_{j}e^{-\lambda_{j}}}{\left(1-e^{-\lambda_{j}}\right)}\right)\right]s_{a}^{n_{a}}\cdots s_{b}^{n_{b}}\\
 & =\sum_{n_{a},\ldots,n_{b}=1}^{\infty}\left[\left(\prod_{i=a}^{b}\frac{\lambda_{i}(1-\rho)\rho^{n_{i}-1}}{n_{i}\left(n_{i}+1\right)}L_{n_{i}-1}^{2}\left(x_{i}\right)\right)f^{w}\left(0,\ldots,0\right)\right]s_{a}^{n_{a}}\cdots s_{b}^{n_{b}},
\end{alignedat}
\end{flalign*}
}{\large\par}

where $x_{i}=-\lambda_{i}\frac{(1-\rho)}{\rho}$, $i\in\{a,\ldots,b\}$.
The result for the second part of the proof follows from this.

In the third and final part of the proof the PMF is derived when all
of the $\left\{ n_{1},\ldots,n_{k}\right\} $ is greater than zero.
Let $\left\{ n_{1},\ldots,n_{k}\right\} \neq\left\{ 0,\ldots,0\right\} $,
 $x_{i}=-\lambda_{i}\frac{(1-\rho)}{\rho}$ and
$z_{i}=\rho s_{i}$ , $i\in\{1,\ldots,k\}$ then by using (\ref{A1}), we
can write the PGF in (\ref{eq:7}) as

{\large{}
\begin{flalign*}
\begin{alignedat}{1} & \psi_{N_{1}^{w},\ldots,N_{k}^{w}}(s_{1},\ldots,s_{k})\\
 & =\sum_{n_{1},\ldots,n_{k}=1}^{\infty}f^{w}\left(n_{1},\ldots,n_{k}\right)s_{1}^{n_{1}}\cdots s_{k}^{n_{k}}\\
 & =\left[\frac{1}{\lambda_{k+1}}\left(e^{\lambda_{k+1}\prod_{i=1}^{k}\left(\frac{(1-\rho)s_{i}}{1-\rho s_{i}}\right)}-1\right)\prod_{i=1}^{k}\frac{1}{\lambda_{i}\left(\frac{(1-\rho)s_{i}}{1-\rho s_{i}}\right)^{2}}\left(e^{\lambda_{i}\left(\frac{(1-\rho)s_{i}}{1-\rho s_{i}}\right)}-1\right)\right]f^{w}\left(0,\ldots,0\right)\\
 & =\left[\frac{1}{\lambda_{k+1}}\left(e^{\lambda_{k+1}\prod_{i=1}^{k}\left(\frac{(1-\rho)}{\rho}\frac{z_{i}}{1-z_{i}}\right)}-1\right)\left(\prod_{i=1}^{k}\frac{1}{\lambda_{i}}\frac{\rho^{2}}{(1-\rho)^{2}}\frac{\left(1-z_{i}\right)^{2}}{z_{i}^{2}}\left(e^{\left(\frac{x_{i}z_{i}}{z_{i}-1}\right)}-1\right)\right)\right]f^{w}\left(0,\ldots,0\right)\\
 & =\left[\left(\prod_{i=1}^{k}\frac{\rho}{\lambda_{i}(1-\rho)}\frac{\left(1-z_{i}\right)}{z_{i}}\left(e^{\left(\frac{x_{i}z_{i}}{z_{i}-1}\right)}-1\right)\right)+\frac{\lambda_{k+1}}{2!}\left(\prod_{i=1}^{k}\frac{1}{\lambda_{i}}\left(e^{\left(\frac{x_{i}z_{i}}{z_{i}-1}\right)}-1\right)\right)\right.\\
 & \left.+\sum_{r=3}^{\infty}\frac{\lambda_{k+1}^{r-1}}{r!}\left(\prod_{i=1}^{k}\frac{(1-\rho)^{r-2}}{\lambda_{i}\rho^{r-2}}\frac{z_{i}^{r-2}}{\left(1-z_{i}\right)^{r-2}}\left(e^{\left(\frac{x_{i}z_{i}}{z_{i}-1}\right)}-1\right)\right)\right]f^{w}\left(0,\ldots,0\right)\\
 & =\psi_{N_{1}^{w},\ldots,N_{k}^{w}}^{\left(1\right)}\left(s_{1},\ldots,s_{k}\right)+\psi_{N_{1}^{w},...,N_{k}^{w}}^{\left(2\right)}\left(s_{1},\ldots,s_{k}\right)+\psi_{N_{1}^{w},...,N_{k}^{w}}^{\left(3\right)}\left(s_{1},\ldots,s_{k}\right).
\end{alignedat}
\end{flalign*}
}{\large\par}

The expression for $\psi_{N_{1},\ldots,N_{k}}^{\left(1\right)}(s_{1},\ldots,s_{k})$
can be rewritten in terms of (\ref{eq:5}) and (\ref{eq:6}), such that

{\large{}
\begin{flalign*}
\begin{alignedat}{1} & \psi_{N_{1}^{w},\ldots,N_{k}^{w}}^{\left(1\right)}\left(s_{1},\ldots,s_{k}\right)\\
 & =\sum_{n_{1},\ldots,n_{k}=1}^{\infty}f^{w\left(1\right)}\left(n_{1},\ldots,n_{k}\right)s_{1}^{n_{1}}\cdots s_{k}^{n_{k}}\\
 & =\left[\prod_{i=1}^{k}\frac{\rho}{\lambda_{i}(1-\rho)}\frac{\left(1-z_{i}\right)}{z_{i}}\left(e^{\left(\frac{x_{i}z_{i}}{z_{i}-1}\right)}-1\right)\right]f^{w}\left(0,\ldots,0\right)\\
 & =\frac{\lambda_{k+1}e^{-\lambda_{k+1}}}{\left(1-e^{-\lambda_{k+1}}\right)}\left[\prod_{i=1}^{k}\frac{e^{-\lambda_{i}}}{\left(1-e^{-\lambda_{i}}\right)}\frac{\left(e^{\lambda_{i}\psi_{1}\left(s_{i}\right)}-1\right)}{\psi_{1}\left(s_{i}\right)}\right]\\
 & =\frac{\lambda_{k+1}e^{-\lambda_{k+1}}}{\left(1-e^{-\lambda_{k+1}}\right)}\prod_{i=1}^{k}\psi_{U_{i}}^{w}\left(s_{i}\right)\\
 & =\sum_{n_{1},\ldots,n_{k}=1}^{\infty}\left[\frac{\lambda_{k+1}e^{-\lambda_{k+1}}}{\left(1-e^{-\lambda_{k+1}}\right)}\prod_{i=1}^{k}\text{Pr}\left(U_{i}^{w}=n_{i}\right)\right]s_{1}^{n_{1}}\cdots s_{k}^{n_{k}}\\
 & =\sum_{n_{1},\ldots,n_{k}=1}^{\infty}\left[\left(\prod_{i=1}^{k}\frac{\lambda_{i}\left(1-\rho\right)\rho^{n_{i}-1}}{n_{i}\left(n_{i}+1\right)}L_{n_{i}-1}^{2}\left(x_{i}\right)\right)f^{w}\left(0,\ldots,0\right)\right]s_{1}^{n_{1}}\cdots s_{k}^{n_{k}}.
\end{alignedat}
\end{flalign*}
}{\large\par}

Using the mathematical results in (\ref{A1}), the generating function for Laguerre polynomials in (\ref{A26}), as well as (\ref{A28}) and  (\ref{eq:A31}), $\psi_{N_{1},\ldots,N_{k}}^{\left(2\right)}(s_{1},\ldots,s_{k})$
and $\psi_{N_{1}^{w},\ldots,N_{k}^{w}}^{\left(3\right)}\left(s_{1},\ldots,s_{k}\right)$
can be rewritten as

{\large{}
\begin{flalign*}
\begin{alignedat}{1} & \psi_{N_{1}^{w},\ldots,N_{k}^{w}}^{\left(2\right)}\left(s_{1},\ldots,s_{k}\right)\\
 & =\sum_{n_{1},\ldots,n_{k}=1}^{\infty}f^{w\left(2\right)}\left(n_{1},\ldots,n_{k}\right)s_{1}^{n_{1}}\cdots s_{k}^{n_{k}}\\
 & =\frac{\lambda_{k+1}}{2!}\left[\prod_{i=1}^{k}\frac{1}{\lambda_{i}}\left(e^{\left(\frac{x_{i}z_{i}}{z_{i}-1}\right)}-1\right)\right]f^{w}\left(0,\ldots,0\right)\\
 & =\frac{\lambda_{k+1}}{2!}\left[\prod_{i=1}^{k}\frac{\left(1-z_{i}\right)}{\lambda_{i}}\sum_{n_{i}=0}^{\infty}\left(L_{n_{i}}^{0}\left(x_{i}\right)-1\right)z_{i}^{n_{i}}\right]f^{w}\left(0,\ldots,0\right)\\
 & =\frac{\lambda_{k+1}}{2!}\left[\prod_{i=1}^{k}\frac{1}{\lambda_{i}}\sum_{n_{i}=1}^{\infty}\left(L_{n_{i}}^{0}\left(x_{i}\right)-L_{n_{i}-1}^{0}\left(x_{i}\right)\right)z_{i}^{n_{i}}\right]f^{w}\left(0,\ldots,0\right)\\
 & =\sum_{n_{1},...,n_{k}=1}^{\infty}\left[\frac{\lambda_{k+1}}{2!}\left(\prod_{i=1}^{k}\frac{\rho^{n_{i}}}{\lambda_{i}}\left(L_{n_{i}}^{0}\left(x_{i}\right)-L_{n_{i}-1}^{0}\left(x_{i}\right)\right)\right)f^{w}\left(0,...,0\right)\right]s_{1}^{n_{1}}\cdots s_{k}^{n_{k}}\\
 & =\sum_{n_{1},\ldots,n_{k}=1}^{\infty}\left[\frac{\lambda_{k+1}}{2!}\left(\prod_{i=1}^{k}\frac{(1-\rho)\rho^{n_{i}-1}}{n_{i}}L_{n_{i}-1}^{1}\left(x_{i}\right)\right)f^{w}\left(0,\ldots,0\right)\right]s_{1}^{n_{1}}\cdots s_{k}^{n_{k}},
\end{alignedat}
\end{flalign*}
}{\large\par}

{\large{}
\begin{flalign*}
\begin{alignedat}{1} & \psi_{N_{1}^{w},\ldots,N_{k}^{w}}^{\left(3\right)}\left(s_{1},\ldots,s_{k}\right)\\
 & =\sum_{n_{1},\ldots,n_{k}=1}^{\infty}f^{w\left(3\right)}\left(n_{1},\ldots,n_{k}\right)s_{1}^{n_{1}}\cdots s_{k}^{n_{k}}\\
 & =\sum_{r=3}^{\infty}\frac{\lambda_{k+1}^{r-1}}{r!}\left[\prod_{i=1}^{k}\frac{(1-\rho)^{r-2}}{\lambda_{i}\rho^{r-2}}\frac{z_{i}^{r-2}}{\left(1-z_{i}\right)^{r-2}}\left(e^{\left(\frac{x_{i}z_{i}}{z_{i}-1}\right)}-1\right)\right]f^{w}\left(0,\ldots,0\right)\\
 & =\sum_{r=3}^{\infty}\frac{\lambda_{k+1}^{r-1}}{r!}\left[\prod_{i=1}^{k}\frac{(1-\rho)^{r-2}z_{i}^{r-2}}{\lambda_{i}\rho^{r-2}}\left(\sum_{n_{i}=0}^{\infty}L_{n_{i}}^{r-3}\left(x_{i}\right)z_{i}^{n_{i}}-\sum_{n_{i}=0}^{\infty}\left(\begin{array}{c}
n_{i}+r-3\\
n_{i}
\end{array}\right)z_{i}^{n_{i}}\right)\right]f^{w}\left(0,\ldots,0\right)\\
 & =\sum_{r=0}^{\infty}\frac{\lambda_{k+1}^{r+2}}{\left(r+3\right)!}\left[\sum_{n_{1},...,n_{k}=0}^{\infty}\left(\prod_{i=1}^{k}\frac{(1-\rho)^{r+1}}{\lambda_{i}\rho^{r+1}}\left(L_{n_{i}}^{r}\left(x_{i}\right)-\left(\begin{array}{c}
n_{i}+r\\
n_{i}
\end{array}\right)\right)\right)z_{1}^{n_{1}+r+1}\cdots z_{k}^{n_{1}+r+1}\right]f^{w}\left(0,\ldots,0\right)\\
 & =\sum_{n_{1},\ldots,n_{k}=1}^{\infty}\left[\sum_{r=1}^{min(n_{1},...,n_{k})}\frac{\lambda_{k+1}^{r+1}}{\left(r+2\right)!}\left(\prod_{i=1}^{k}\frac{(1-\rho)^{r}\rho^{n_{i}-r}}{\lambda_{i}}\left(L_{n_{i}-r}^{r-1}\left(x_{i}\right)-\left(\begin{array}{c}
n_{i}-1\\
n_{i}-r
\end{array}\right)\right)\right)f^{w}\left(0,\ldots,0\right)\right]s_{1}^{n_{1}}\cdots s_{k}^{n_{k}}.
\end{alignedat}
\end{flalign*}
}{\large\par}

It follows that

{\large{}
\begin{flalign*}
\begin{alignedat}{1}f^{w}\left(n_{1},\ldots,n_{k}\right) & =f^{w\left(1\right)}\left(n_{1},\ldots,n_{k}\right)+f^{w\left(2\right)}\left(n_{1},\ldots,n_{k}\right)+f^{w\left(3\right)}\left(n_{1},\ldots,n_{k}\right)\end{alignedat}
.
\end{flalign*}
}{\large\par}

\end{proof}

\subsection{{Method of moments estimates}}

Let $\left(N_{1}^{w},\ldots,N_{k}^{w}\right)\sim WMPA_{I}\left(\lambda_{1},\ldots,\lambda_{k},\lambda_{k+1},\rho\right)$
and consider a random sample of size $m$ from this distribution.
The observed values can then be given as $\left(n_{1\ell},\ldots,n_{k\ell}\right)$
where $\ell\in\{1,\ldots,m\}$. In addition, for $i,j\in\{1,\ldots,k\}$,
$i\neq j$, let

{\large{}
\begin{flalign*}
\begin{alignedat}{1}\overline{n}_{i}^{w} & =\frac{1}{m}\sum_{\ell=1}^{m}n_{i\ell}\\
s_{ij}^{w} & =\frac{1}{\left(m-1\right)}\sum_{\ell=1}^{m}\left(n_{i\ell}-\overline{n}_{i}\right)\left(n_{j\ell}-\overline{n}_{j}\right)\\
s_{i}^{2^{w}} & =\frac{1}{\left(m-1\right)}\sum_{\ell=1}^{m}\left(n_{i\ell}-\overline{n}_{i}\right)^{2}.
\end{alignedat}
\end{flalign*}
}{\large\par}

Using these together with the expressions in (\ref{eq:8}) for $\text{E}\left(N_{i}^{w}\right)=\mu_{i}$,
$\text{Cov}\left(N_{i}^{w},N_{j}^{w}\right)=\mu_{ij}-\mu_{i}\mu_{j}$
and $\text{Var}\left(N_{i}^{w}\right)=\sigma_{i}^{2}$, we equate
the first sample moments $\overline{n}_{i}^{w}$ as well as $s_{ij}^{w}$
and $s_{i}^{2^{w}}$ to $\mu_{i}$, $\mu_{ij}-\mu_{i}\mu_{j}$ and
$\sigma_{i}^{2}$ respectively. Since we cannot find explicit expressions
for the MoM estimates of the parameters, we will use the Newton-Rhapson
algorithm to obtain the parameter estimates of $\lambda_{1}$ and
$\lambda_{k+1}$. If $\hat{\lambda}_{1}$ and $\hat{\lambda}_{k+1}$
are estimates for $\lambda_{1}$ and $\lambda_{k+1}$ respectively,
then an estimate for $\rho$ is

{\large{}
\begin{flalign}
\begin{alignedat}{1}\hat{\rho} & =1-\sqrt{\frac{k!}{\left(k-2\right)!}\frac{\left(\hat{\lambda}_{k+1}\left(1-e^{-\hat{\lambda}_{k+1}}\right)-\hat{\lambda}_{k+1}^{2}e^{-\hat{\lambda}_{k+1}}\right)}{\left(1-e^{-\hat{\lambda}_{k+1}}\right)^{2}\sum_{i,j=1,i\neq j}^{k}s_{ij}^{w}}}\end{alignedat}
.\label{eq:12}
\end{flalign}
}{\large\par}

Substituting the estimate for $\rho$ in (\ref{eq:12}) into the expressions
for $\hat{\mu}_{1}$ and $\hat{\sigma}_{1}^{2}$, we define the function
$g\left(\hat{\lambda}_{1},\hat{\lambda}_{k+1}\right)$ as

{\large{}
\begin{flalign*}
\begin{alignedat}{1}g\left(\hat{\lambda}_{1},\hat{\lambda}_{k+1}\right) & =\left[\begin{array}{c}
g_{1}\left(\hat{\lambda}_{1},\hat{\lambda}_{k+1}\right)\\
g_{2}\left(\hat{\lambda}_{1},\hat{\lambda}_{k+1}\right)
\end{array}\right]\\
 & =\left[\begin{array}{c}
\hat{\mu}_{1}-\overline{n}_{1}^{w}\\
\hat{\sigma}_{1}^{2}-s_{1}^{2^{w}}
\end{array}\right]\\
 & =\left[\begin{array}{c}
\frac{1}{\left(1-\hat{\rho}\right)}\left[\frac{\hat{\lambda}_{1}}{\left(1-e^{-\hat{\lambda}_{1}}\right)}+\frac{\hat{\lambda}_{k+1}}{\left(1-e^{-\hat{\lambda}_{k+1}}\right)}-2\right]-\overline{n}_{1}^{w}\\
\frac{\left(1+\hat{\rho}\right)}{\left(1-\hat{\rho}\right)^{2}}\left[\frac{\hat{\lambda}_{1}}{\left(1-e^{-\hat{\lambda}_{1}}\right)}+\frac{\hat{\lambda}_{k+1}}{\left(1-e^{-\hat{\lambda}_{k+1}}\right)}-2\right]-\frac{1}{\left(1-\hat{\rho}\right)^{2}}\left[\frac{\hat{\lambda}_{1}^{2}e^{-\hat{\lambda}_{1}}}{\left(1-e^{-\hat{\lambda}_{1}}\right)^{2}}+\frac{\hat{\lambda}_{k+1}^{2}e^{-\hat{\lambda}_{k+1}}}{\left(1-e^{-\hat{\lambda}_{k+1}}\right)^{2}}-2\right]-s_{1}^{2^{w}}
\end{array}\right].
\end{alignedat}
\end{flalign*}
}{\large\par}

We can now find the expression for the Jacobian matrix

{\large{}
\begin{flalign*}
\begin{alignedat}{1}J_{g}\left(\hat{\lambda}_{1},\hat{\lambda}_{k+1}\right) & =\left[\begin{array}{cc}
\frac{\partial g_{1}}{\partial\hat{\lambda}_{1}} & \frac{\partial g_{1}}{\partial\hat{\lambda}_{k+1}}\\
\frac{\partial g_{2}}{\partial\hat{\lambda}_{1}} & \frac{\partial g_{2}}{\partial\hat{\lambda}_{k+1}}
\end{array}\right]\end{alignedat}
\end{flalign*}
}{\large\par}

and use the Newton-Rhapson algorithm to solve for the parameter estimates
for $\lambda_{1}$ and $\lambda_{k+1}$

{\large{}
\begin{flalign*}
\begin{alignedat}{2}\left[\begin{array}{c}
\hat{\lambda}_{1}^{\left(t+1\right)}\\
\hat{\lambda}_{k+1}^{\left(t+1\right)}
\end{array}\right] & = & \left[\begin{array}{c}
\hat{\lambda}_{1}^{\left(t\right)}\\
\hat{\lambda}_{k+1}^{\left(t\right)}
\end{array}\right]-\left[J_{g}\left(\hat{\lambda}_{1}^{\left(t\right)},\hat{\lambda}_{k+1}^{\left(t\right)}\right)\right]^{-1}g\left(\hat{\lambda}_{1}^{\left(t\right)},\hat{\lambda}_{k+1}^{\left(t\right)}\right)\end{alignedat}
\end{flalign*}
}{\large\par}

for $t\geq1$. We will use the MoM estimates from the Type I MPA as
initial values of the parameters $\hat{\lambda}_{1}^{\left(0\right)}$and
$\hat{\lambda}_{k+1}^{\left(0\right)}$. The iterative process is
repeated until we reach a given tolerance level $\epsilon$ between
the $t^{th}$ and $\left(t+1\right)^{th}$ iterative values or until
a specified maximum number of iterations is reached. Lastly, we can
obtain an estimate for $\rho$ by substituting the parameter estimates
obtained through the iterative process for $\lambda_{1}$ and $\lambda_{k+1}$
into (\ref{eq:12}). We are also now able to find estimates for the
rest of the $\lambda_{i}$s, $i\in\{2,\ldots,k\}$

{\large{}
\begin{flalign*}
\begin{alignedat}{1}\hat{\lambda_{i}} & =(1-\rho)\left[\overline{n}_{i}^{w}-\frac{\hat{\lambda}_{k+1}}{(1-\hat{\rho})\left(1-e^{-\hat{\lambda}_{k+1}}\right)}+\frac{2}{(1-\hat{\rho})}\right]+\frac{(1-\hat{\rho})\left(s_{i}^{2^{w}}-\sum_{j=1,j\neq i}^{k}\frac{s_{ij}^{w}}{\left(k-1\right)}\right)-\hat{\rho}\overline{n}_{i}^{w}}{\left[\overline{n}_{i}^{w}-\frac{\hat{\lambda}_{k+1}}{(1-\hat{\rho})\left(1-e^{-\hat{\lambda}_{k+1}}\right)}+\frac{2}{(1-\hat{\rho})}\right]}-1.\end{alignedat}
\end{flalign*}
}{\large\par}

\section{Maximum likelihood estimation}\label{sec:4}

A situation can arise where we are unable to find MoM estimates
for the parameters of these distributions. In this case we can calculate
the MLEs using a Newton-Raphson algorithm
\cite{balakrishnan2017likelihood}. The likelihood and log-likelihood functions can be written
as

{\large{}
\[
\begin{alignedat}{1}L\left(\lambda_{1},...,\lambda_{k},\lambda_{k+1},\rho|n_{1},...,n_{k}\right) & =\prod_{\ell=1}^{m}f\left(N_{1\ell,},...,N_{k\ell}\right),\end{alignedat}
\]
}{\large\par}

{\large{}
\[
\begin{alignedat}{1}l\left(\lambda_{1},...,\lambda_{k},\lambda_{k+1},\rho|n_{1},...,n_{k}\right) & =ln\left(L\left(\lambda_{1},...,\lambda_{k},\lambda_{k+1},\rho|n_{1},...,n_{k}\right)\right)\\
 & =\sum_{\ell=1}^{m}ln\left(f\left(N_{1\ell,},...,N_{k\ell}\right)\right)
\end{alignedat}
\]
}{\large\par}

respectively, where $m$ is the sample size. The Jacobian matrix is
then

{\large{}
\[
\begin{alignedat}{1} & J\left(\lambda_{1},...,\lambda_{k},\lambda_{k+1},\rho\right)\\
 & =\left[\begin{array}{ccc}
\sum_{\ell=1}^{m}\left[\frac{\frac{\partial^{2}}{\partial\lambda_{1}^{2}}f\left(n_{1},...,n_{k}\right)}{f\left(n_{1},...,n_{k}\right)}-\frac{\frac{\partial}{\partial\lambda_{1}}f^{2}\left(n_{1},...,n_{k}\right)}{f\left(n_{1},...,n_{k}\right)^{2}}\right] & \cdots & \sum_{\ell=1}^{m}\left[\frac{\frac{\partial^{2}}{\partial\lambda_{1}\rho}f\left(n_{1},...,n_{k}\right)}{f\left(n_{1},...,n_{k}\right)}-\frac{\frac{\partial}{\partial\lambda_{1}}f\left(n_{1},...,n_{k}\right)\frac{\partial}{\partial\rho}f\left(n_{1},...,n_{k}\right)}{f\left(n_{1},...,n_{k}\right)^{2}}\right]\\
\vdots & \ddots & \vdots\\
\sum_{\ell=1}^{m}\left[\frac{\frac{\partial^{2}}{\partial\lambda_{1}\rho}f\left(n_{1},...,n_{k}\right)}{f\left(n_{1},...,n_{k}\right)}-\frac{\frac{\partial}{\partial\lambda_{1}}f\left(n_{1},...,n_{k}\right)\frac{\partial}{\partial\rho}f\left(n_{1},...,n_{k}\right)}{f\left(n_{1},...,n_{k}\right)^{2}}\right] & \cdots & \sum_{\ell=1}^{m}\left[\frac{\frac{\partial^{2}}{\partial\rho^{2}}f\left(n_{1},...,n_{k}\right)}{f\left(n_{1},...,n_{k}\right)}-\frac{\frac{\partial}{\partial\rho}f^{2}\left(n_{1},...,n_{k}\right)}{f\left(n_{1},...,n_{k}\right)^{2}}\right]
\end{array}\right].
\end{alignedat}
\]
}{\large\par}

The MLEs can then be found using the following
Newton-Raphson iterative process

{\large{}
\[
\begin{alignedat}{1}\left[\begin{array}{c}
\lambda_{1}^{\left(t\right)}\\
\vdots\\
\lambda_{k}^{\left(t\right)}\\
\lambda_{k+1}^{\left(t\right)}\\
\rho^{\left(t\right)}
\end{array}\right] & =\left[\begin{array}{c}
\lambda_{1}^{\left(t-1\right)}\\
\vdots\\
\lambda_{k}^{\left(t\right)}\\
\lambda_{k+1}^{\left(t\right)}\\
\rho^{\left(t\right)}
\end{array}\right]-J^{-1}\left[\begin{array}{c}
\lambda_{1}^{\left(t-1\right)}\\
\vdots\\
\lambda_{k}^{\left(t\right)}\\
\lambda_{k+1}^{\left(t\right)}\\
\rho^{\left(t\right)}
\end{array}\right]F\left[\begin{array}{c}
\lambda_{1}^{\left(t-1\right)}\\
\vdots\\
\lambda_{k}^{\left(t\right)}\\
\lambda_{k+1}^{\left(t\right)}\\
\rho^{\left(t\right)}
\end{array}\right]\end{alignedat}
\]
}{\large\par}

for $t\geq1$ and where we will use the MoM estimates as initial
values of the parameters $\lambda_{1}^{\left(0\right)},...,\lambda_{k}^{\left(0\right)},\lambda_{k+1}^{\left(0\right)}$
and $\rho^{\left(0\right)}$. The iterative process is repeated until
we reach a given tolerance level $\epsilon$ between the $t^{th}$
and $\left(t+1\right)^{th}$ iterative values or until a specified maximum
number of iterations is reached \cite{balakrishnan2017likelihood}.

It is worth noting that in \cite{balakrishnan2017likelihood} the PMFs and therefore log-likelihood functions are recursive. With this alternative approach to calculate the PMF of the Type I MPA that was extended to the Type I WMPA, these functions are much simpler than the recursive expressions. This implies a simpler structure and method for computation of the MLEs.

\section{Generalized dispersion index}\label{sec:5}

In order to understand the dispersion of the Type I MPA and Type I WMPA, we utilize GDI for multivariate distributions proposed in \cite{kokonendji2018fisher}. Let $\textbf{N}=\left(N_{1},\ldots,N_{k}\right)^\text{T}$, be a
nondegenerate count $k$-variate random vector on $\mathbb{N}^{k}$,
$k\geq1$. Then $\sqrt{\text{E}\textbf{(N)}}=\left(\sqrt{\text{E}(N_{1})},\ldots,\sqrt{\text{E}(N_{k})}\right)^{\text{T}}$is
the elementwise square root of the mean vector of $\textbf{N}$ and $\text{Cov}\textbf{(N)}=\left(\text{Cov}\left(N_{i},N_{j}\right)\right)_{i,j\in\{1,\ldots,k\}}$
denotes the covariance matrix of $\textbf{N}$ which is a $k\times k$ symmetric
matrix with entries $\text{Cov}\left(N_{i},N_{j}\right)$ such that
$\text{Cov}\left(N_{i},N_{i}\right)=\text{Var}\left(N_{i}\right)$
is the variance of $N_{i}$. It follows that the GDI of $\textbf{N}$ is defined as

{\large{}
\[
\begin{alignedat}{1}\text{GDI}\left(\textbf{N}\right) & =\frac{\sqrt{\text{E}\textbf{(N)}}^{\text{T}}\left(\text{Cov}\textbf{(N)}\right)\sqrt{\text{E}\textbf{(N)}}}{\text{E}\textbf{(N)}^{\text{T}}\text{E}\textbf{(N)}}\end{alignedat}
.
\]
}{\large\par}

For the uncorrelated multivariate Poisson distribution, $\text{GDI}\left(\textbf{N}\right)=1$. The GDI for the Type I BPA is also given in \cite{kokonendji2018fisher}, illustrating that it is always overdispersed, as

{\large{}
\[
\begin{alignedat}{1}\text{GDI}\left(\textbf{N}\right) & =1+2\frac{\rho\left(\lambda_{1}+\lambda_{3}\right)^{2}+\rho\left(\lambda_{2}+\lambda_{3}\right)^{2}+\lambda_{3}\sqrt{\left(\lambda_{1}+\lambda_{3}\right)\left(\lambda_{2}+\lambda_{3}\right)}}{\left(1-\rho\right)\left(\lambda_{1}+\lambda_{3}\right)^{2}+\left(1-\rho\right)\left(\lambda_{2}+\lambda_{3}\right)^{2}}\\
 & =1+\frac{2\rho}{\left(1-\rho\right)}+\frac{2\lambda_{3}\sqrt{\left(\lambda_{1}+\lambda_{3}\right)\left(\lambda_{2}+\lambda_{3}\right)}}{\left(1-\rho\right)\left[\left(\lambda_{1}+\lambda_{3}\right)^{2}+\left(\lambda_{2}+\lambda_{3}\right)^{2}\right]}>1,
\end{alignedat}
\]
}{\large\par}

where $\textbf{N}=\left(N_{1},N_{2}\right)^{\text{T}}$. We can easily extend
this for the Type I MPA distribution such that

{\large{}
\[
\begin{alignedat}{1}\text{GDI}\left(\textbf{N}\right) & =1+\frac{2\rho}{\left(1-\rho\right)}+\frac{2\lambda_{k+1}\sqrt{\prod_{i=1}^{k}\left(\lambda_{i}+\lambda_{k+1}\right)}}{\left(1-\rho\right)\sum_{i=1}^{k}\left(\lambda_{i}+\lambda_{k+1}\right)^{2}}>1,\end{alignedat}
\]
}{\large\par}

where $\textbf{N}=\left(N_{1},\ldots,N_{k}\right)^{\text{T}}$ and $\ensuremath{i\in\{1,\ldots,k\}}$. This illustrates that the Type I MPA distribution is always overdispersed relative
to the uncorrelated multivariate Poisson distribution.

The overdispersion of the Type I WMPA
distribution is more difficult to visualize mathematically, therefore
we represent the GDI of this distribution in Figure 1, using the GDI
of the Type I MPA as a reference.

For illustration purposes, we will present the GDI of the bivariate
distributions varying the parameters of $\lambda_{1}$, $\lambda_{3}$
and $\rho$ one at a time, while keeping the other parameters constant. The GDI will behave in the same way for varying values of $\lambda_{2}$ as for $\lambda_{1}$.

We can see that the Type I WMPA is not only overdispersed but also overdispersed with respect to the Type I MPA.

\begin{figure}[H]
\noindent\fbox{\begin{minipage}[t]{1\columnwidth \fboxsep \fboxrule}%
\[
\begin{alignedat}{2}\\
\includegraphics[width=0.3\linewidth]{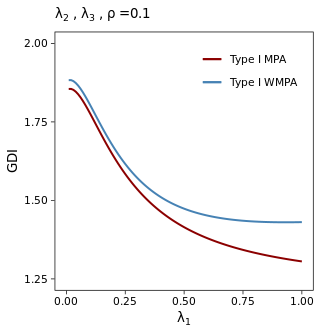}    
\includegraphics[width=0.3\linewidth]{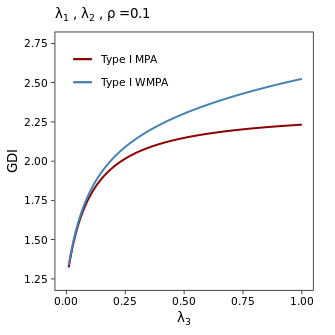}
\includegraphics[width=0.3\linewidth]{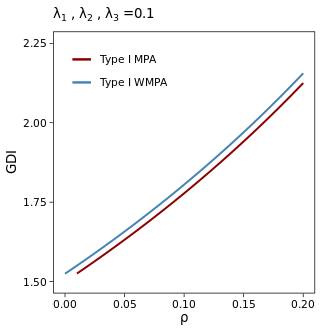}
\\
\includegraphics[width=0.3\linewidth]{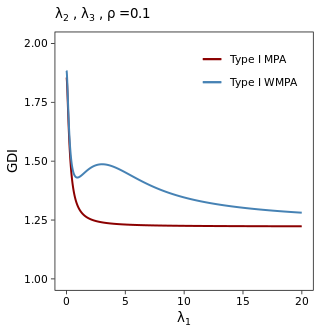}
\includegraphics[width=0.3\linewidth]{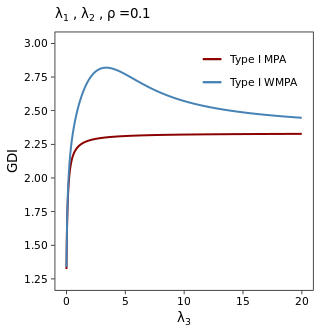}    
\includegraphics[width=0.3\linewidth]{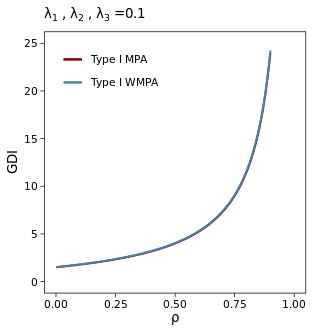}
\end{alignedat}
\]
\end{minipage}}

\caption{Generalized dispersion index of Type I MPA and Type I WMPA}

\end{figure}

\section{Simulation Study}\label{sec:6}

A simulation study was carried out in \cite{balakrishnan2017likelihood} to evaluate the performance of the MLEs and MoMs for the Type I bivariate Pólya-Aeppli distribution under different parameter settings. The simulations produced 1000 replications of a sample series under different parameter settings.

Similarly we have carried out a simulation study in the bivariate case of 1000 replications of a sample series when we were able to find both MoM estimates as well as MLEs. We chose the values for the parameters as follows $\lambda_{1}=0.6$, $\lambda_{2}=0.6$, $\lambda_{3}=0.3$, $\rho=0.1$ in order to evaluate the performance of the estimation methods across various sample sizes for the Type I MPA and Type I WMPA distributions.

Since both distibutions use the Newton-Raphson algorithm to calculate the MLEs, we set the tolerance level $\epsilon=0.01$ between the $t^{th}$ and $\left(t+1\right)^{th}$ iterative values for each of the parameters, in order to have a good balance between accuracy and computation time.

Since the MoM estimates for the Type I WMPA distribution also uses the Newton-Raphson algorithm, we set the tolerance level $\epsilon=0.001$ between the $t^{th}$ and $\left(t+1\right)^{th}$ iterative values for each of the parameters. We set the value for $\epsilon=0.001$ smaller for the MoM estimates, since we use these as initial values for the MLE of the parameters. The enhanced precision in the MoM similarly enhances the precision of the MLEs. The smaller $\epsilon$ in the case of the MoM estimates also doesn't noticeably effect computation time.

In Tables 1 and 2 we present the mean, bias and MSE of the 2 distributions. We note that in both cases across all sample sizes, the MLEs tend to perform better with smaller MSEs.

\begin{table}[H]
		\caption{Moment Estimates for $\lambda_{1}=0.6$, $\lambda_{2}=0.6$, $\lambda_{3}=0.3$, $\rho=0.1$}
		\setlength{\tabcolsep}{4pt}
		\begin{tabular}{lcccc}
			\noalign{\smallskip}\hline\noalign{\smallskip}
            & \multicolumn{4}{c}{Type I MPA (Mean, Bias, MSE)}\\			
            \noalign{\smallskip}\cline{2-5}\noalign{\smallskip} 
            s & $\lambda_{1}$ & $\lambda_{2}$ & $\lambda_{3}$ & $\rho$\\
			\noalign{\smallskip}\hline\noalign{\smallskip}
            $50$ & $\left(0.6079,0.0079,0.0337\right)$ & $\left(0.5980,-0.0020,0.0333\right)$ & $\left(0.3042,0.0042,0.0193\right)$ & $\left(0.1039,0.0039,0.0039\right)$ \\
			$100$ &$\left(0.6084,0.0084,0.0187\right)$ & $\left(0.6090,0.0090,0.0190\right)$ & $\left(0.3004,0.0004,0.0108\right)$ & $\left(0.0940,-0.0060,0.0027\right)$\\
			$200$ & $\left(0.6052,0.0052,0.0100\right)$ & $\left(0.6068,0.0068,0.0105\right)$ & $\left(0.2962,-0.0038,0.0055\right)$ & $\left(0.0959,-0.0041,0.0016\right)$\\
			$500$ & $\left(0.6036,0.0036,0.0042\right)$ & $\left(0.6014,0.0014,0.0043\right)$ & $\left(0.2987,-0.0013,0.0024\right)$ & $\left(0.0979,-0.0021,0.0006\right)$\\
			\noalign{\smallskip}\hline\noalign{\smallskip} 
            & \multicolumn{4}{c}{Type I WMPA (Mean, Bias, MSE)}\\	
            \noalign{\smallskip}\cline{2-5}\noalign{\smallskip} 
            s & $\lambda_{1}$ & $\lambda_{2}$ & $\lambda_{3}$ & $\rho$\\
            \noalign{\smallskip}\hline\noalign{\smallskip}
            $50$ & $\left(0.5884,-0.0116,0.0411\right)$ & $\left(0.5304,-0.0696,0.0496\right)$ & $\left(0.2985,-0.0015,0.0234\right)$ & $\left(0.1416,0.0416,0.0104\right)$\\
			$100$ & $\left(0.5972,-0.0028,0.0213\right)$ & $\left(0.5442,-0.0558,0.0302\right)$ & $\left(0.284,-0.016,0.0157\right)$ & $\left(0.1374,0.0374,0.0077\right)$ \\
			$200$ & $\left(0.6075,0.0075,0.0121\right)$ & $\left(0.5709,-0.0291,0.0187\right)$ & $\left(0.2854,-0.0146,0.0089\right)$ & $\left(0.1282,0.0282,0.0057\right)$ \\
			$500$ & $\left(0.6128,0.0128,0.0046\right)$ & $\left(0.5691,-0.0309,0.0111\right)$ & $\left(0.2800,-0.0200,0.0043\right)$ & $\left(0.1300,0.0300,0.0050\right)$\\
            \noalign{\smallskip}\hline\noalign{\smallskip} 
		\end{tabular}
\end{table}

\begin{table}[H]
		\caption{Maximum Likelihood Estimates for $\lambda_{1}=0.6$, $\lambda_{2}=0.6$, $\lambda_{3}=0.3$, $\rho=0.1$}
		\setlength{\tabcolsep}{4pt}
		\begin{tabular}{lcccc}
			\noalign{\smallskip}\hline\noalign{\smallskip}
            & \multicolumn{4}{c}{Type I MPA (Mean, Bias, MSE)}\\			
            \noalign{\smallskip}\cline{2-5}\noalign{\smallskip} 
            s & $\lambda_{1}$ & $\lambda_{2}$ & $\lambda_{3}$ & $\rho$\\
			\noalign{\smallskip}\hline\noalign{\smallskip}
            $50$ & $\left(0.6274,0.0274,0.0314\right)$ & $\left(0.6184,0.0184,0.0297\right)$ & $\left(0.2974,-0.0026,0.0159\right)$ & $\left(0.0913,-0.0087,0.0041\right)$ \\
			$100$ & $\left(0.6227,0.0227,0.0173\right)$ & $\left(0.6237,0.0237,0.0181\right)$ & $\left(0.2973,-0.0027,0.0090\right)$ & $\left(0.0827,-0.0173,0.0032\right)$ \\
			$200$ & $\left(0.6197,0.0197,0.0101\right)$ & $\left(0.6212,0.0212,0.0101\right)$ & $\left(0.2920,-0.0080,0.0046\right)$ & $\left(0.0857,-0.0143,0.0022\right)$ \\
			$500$ & $\left(0.6147,0.0147,0.0047\right)$ & $\left(0.6121,0.0121,0.0047\right)$ & $\left(0.2967,-0.0033,0.0020\right)$ & $\left(0.0892,-0.0108,0.0013\right)$\\
			\noalign{\smallskip}\hline\noalign{\smallskip} 
            & \multicolumn{4}{c}{Type I WMPA (Mean, Bias, MSE)}\\	
            \noalign{\smallskip}\cline{2-5}\noalign{\smallskip} 
            s & $\lambda_{1}$ & $\lambda_{2}$ & $\lambda_{3}$ & $\rho$\\
            \noalign{\smallskip}\hline\noalign{\smallskip}
            $50$ & $\left(0.5740,-0.0260,0.0386\right)$ & $\left(0.5859,-0.0141,0.0365\right)$ & $\left(0.3227,0.0227,0.0175\right)$ & $\left(0.0985,-0.0015,0.0048\right)$\\
			$100$ & $\left(0.5921,-0.0079,0.0195\right)$ & $\left(0.5945,-0.0055,0.0204\right)$ & $\left(0.3051,0.0051,0.0109\right)$ & $\left(0.091,-0.009,0.0031\right)$ \\
			$200$ & $\left(0.6056,0.0056,0.0116\right)$ & $\left(0.6102,0.0102,0.0123\right)$ & $\left(0.3010,0.0010,0.0053\right)$ & $\left(0.0890,-0.0110,0.0024\right)$ \\
			$500$ & $\left(0.6128,0.0128,0.0054\right)$ & $\left(0.6134,0.0134,0.0055\right)$ & $\left(0.2967,-0.0033,0.0022\right)$ & $\left(0.0876,-0.0124,0.0014\right)$\\
            \noalign{\smallskip}\hline\noalign{\smallskip} 
		\end{tabular}
\end{table}

The success rates, in other words the proportion of times we are able to find MoM and MLE estimates was also evaluated in \cite{balakrishnan2017likelihood}. We find that both our distributions perform similarly as was outlined for the Type I BPA.

\section{Application to zero-inflated and zero-and-one inflated over-dispersed count data}\label{sec:7}

Many different models have been proposed for modelling zero-inflated and zero-and-one inflated count data. The literature in the univariate case is extensive, while multivariate distributions to model these kinds of data is far less advanced. As these types of data occur frequently in manufacturing processes, healthcare data, number of accidents, etc. it is increasingly important to have multivariate distributions that can performs well in these situations.

Some of the multivariate models that have been proposed for modelling zero-inflated data are the multivariate zero-inflated Poisson model, a mixture of $m+2$ components of m-dimensional discrete distributions \cite{li1999multivariate}. Using a similar structure, \cite{dong2014multivariate} proposed a multivariate zero-inflated negative binomial model (MINB). In addition, zero-inflated versions for the multivariate Poisson model (MP) was considered in \cite{bermudez2011bayesian}. A new Type I multivariate zero-inflated Poisson distribution (Type 1 MZIP) was introduced in \cite{liu2015type} as an extension of the univariate zero-inflated Poisson (ZIP). Inspired by \cite{liu2015type}, \cite{zhang2022new} developed a multivariate zero-inflated hurdle model.

Multivariate distributions to model count-data with a large number of both zero's and one's have also been proposed. A multivariate zero-and-one inflated Poisson (MZOIP) distribution was introduced in \cite{zhang2020multivariate} as an extension of the zero-and-one-inflated Poisson (ZOIP) distribution in \cite{melkersson1999visiting} and \cite{zhang2016properties}. The bivariate Poisson (BP) and diagonal inflated bivariate Poisson (DIBP) \cite{karlis2005bivariate} have also been proposed for modelling zero-and-one inflated bivariate data.

In the examples below, we will consider the fit of both the Type I MPA and Type I WMPA to overdispersed zero-inflated and zero-and-one-inflated datasets. Considering the characteristics of these two distributions, we anticipate that they will effectively fit overdispersed count data. Moreover, we will observe that they are advantageous in modeling overdispersed zero-inflated and zero-and-one inflated count data, sometimes even more so than the currently available distributions.

\subsection{Example 1}

In this example we will consider a bivariate dataset first used in \cite{cameron2013regression}, concerning the demand for health care in Australia from the Australian health survey for 1977-1978. 

This dataset, without covariates was used in \cite{zhang2020multivariate} to illustrate and compare the fit of the multivariate zero-and-one inflated Poisson distribution (MZOIP) with the Type I multivariate zero-inflated Poisson distribution (Type I MZIP) \cite{liu2015type}, as well as the bivariate Poisson (BP) and diagonal inflated bivariate Poisson (DIBP) \cite{karlis2005bivariate}.

In Table 3, we illustrate the observed frequencies of the dataset as well as the expected frequencies of both the Type I  MPA and Type I WMPA distribution, where $N_1$ denotes the number of consultations with a doctor or a specialist and $N_2$ denotes the total number of prescribed medications used in the past two days.

\begin{table}[H]
		\caption{Observed and expected frequencies of the health care utilization data in the Australian health survey
        for 1977-1978 \cite{cameron2013regression}}
		\setlength{\tabcolsep}{5pt}
		\begin{tabular}{lcccccccccccc}
			\noalign{\smallskip}\hline\noalign{\smallskip}  
			& $N_1/N_2$ & 0 & 1 & 2 & 3 & 4 & 5 & 6 & 7 & 8 & Total\\
			\noalign{\smallskip}\hline\noalign{\smallskip} 
			Observed & 0  & 2789 & 726 & 307 & 171 & 76 & 32 & 16 & 15 & 9 & 4141\\
			Type I MPA &  & 2754.52 & 745.03 & 359.98 & 169.39 & 78.07 & 35.38 & 15.81 & 6.98 & 3.05 & 4169.53\\
			Type I WMPA & & 2798.26 & 708.21 & 357.75 & 175.67 & 84.33 & 39.73 & 18.42 & 8.43 & 3.81 & 4194.61\\
			Observed & 1 & 224 & 212 & 149 & 85 & 50 & 35 & 13 & 5 & 9 & 782\\
			Type I MPA & & 167.03 & 192.39 & 112.87 & 61.19 & 31.50 & 15.63 & 7.54 & 3.56 & 1.65 & 594.11\\
			Type I WMPA & & 163.30 & 189.88 & 108.45 & 58.70 & 30.55 & 15.42 & 7.59 & 3.66 & 1.74 & 579.29\\
			Observed & 2 & 49 & 34 & 38 & 11 & 23 & 7 & 5 & 3 & 4 & 174\\
			Type I MPA & & 63.18 & 77.24 & 49.39 & 28.49 & 15.41 & 7.97 & 3.99 & 1.94 & 0.93 & 248.96\\
			Type I WMPA & & 61.29 & 74.16 & 47.66 & 27.84 & 15.32 & 8.08 & 4.12 & 2.05 & 1.00 & 241.52\\
   		Observed & 3  & 8 & 10 & 6 & 2 & 1 & 1 & 2 & 0 & 0 & 30\\
			Type I MPA & & 23.85 & 30.75 & 21.12 & 12.82 & 7.22 & 3.86 & 1.99 & 0.99 & 0.49 & 103.32\\
			Type I WMPA & & 22.94 & 28.79 & 20.33 & 12.64 & 7.28 & 3.99 & 2.10 & 1.08 & 0.54 & 99.69\\
   		Observed & 4 & 8 & 8 & 2 & 2 & 3 & 1 & 0 & 0 & 0 & 24\\
			Type I MPA & & 8.98 & 12.16 & 8.88 & 5.62 & 3.28 & 1.80 & 0.95 & 0.49 & 0.24 & 42.52\\
			Type I WMPA & & 8.57 & 11.12 & 8.48 & 5.56 & 3.34 & 1.89 & 1.02 & 0.54 & 0.27 & 40.79\\
   		Observed & 5 & 3 & 3 & 2 & 0 & 1 & 0 & 0 & 0 & 0 & 9\\
			Type I MPA & & 3.38 & 4.78 & 3.68 & 2.42 & 1.45 & 0.82 & 0.44 & 0.23 & 0.12 & 17.37\\
			Type I WMPA &  & 3.19 & 4.27 & 3.48 & 2.39 & 1.49 & 0.87 & 0.48 & 0.26 & 0.14 & 16.57\\
   		Observed & 6 & 2 & 0 & 1 & 3 & 1 & 2 & 2 & 0 &  1 & 12\\
			Type I MPA & & 1.27 & 1.87 & 1.51 & 1.02 & 0.63 & 0.36 & 0.20 & 0.11 & 0.05 & 7.05\\
			Type I WMPA &  & 1.19 & 1.64 & 1.41 & 1.01 & 0.65 & 0.39 & 0.22 & 0.12 & 0.06 & 6.69\\
   		Observed & 7 & 1 & 0 & 3 & 2 & 1 & 2 & 1 & 0 &  2 & 12\\
			Type I MPA & & 0.48 & 0.73 & 0.61 & 0.43 & 0.27 & 0.16 & 0.09 & 0.05 & 0.03 & 2.85\\
			Type I WMPA &  & 0.44 & 0.62 & 0.56 & 0.42 & 0.28 & 0.17 & 0.10 & 0.06 & 0.03 & 2.68\\
   		Observed & 8 & 1 & 1 & 1 & 0 & 1 & 0 & 1 & 0 & 0 & 5\\
			Type I MPA & & 0.18 & 0.28 & 0.25 & 0.18 & 0.11 & 0.07 & 0.04 & 0.02 & 0.01 & 1.14\\
			Type I WMPA &  & 0.16 & 0.24 & 0.22 & 0.17 & 0.12 & 0.07 & 0.04 & 0.02 & 0.01 & 1.05\\
   		Observed & 9 & 0 & 0 & 0 & 0 & 0 & 0 & 0 & 0 & 1 & 1\\
			Type I MPA & & 0.07 & 0.11 & 0.10 & 0.07 & 0.05 & 0.03 & 0.02 & 0.01 & 0.01 & 0.46\\
			Type I WMPA &  & 0.06 & 0.09 & 0.09 & 0.07 & 0.05 & 0.03 & 0.02 & 0.01 & 0.01 & 0.43\\
			Observed & Total  & 3085 & 994 & 509 & 276 & 157 & 80 & 40 & 23 & 26 & 5190\\
			Type I MPA & & 3022.92 & 1065.33 & 558.37 & 281.63 & 138.00 & 66.09 & 31.07 & 14.38 & 6.57 & 5184.36\\
			Type I WMPA &  & 3059.40 & 1019.02 & 548.43 & 284.47 & 143.41 & 70.64 & 34.11 & 16.23 & 7.61 & 5183.32\\
			\noalign{\smallskip}\hline\noalign{\smallskip} 
		\end{tabular}
\end{table}

In Table 4 we compare estimation methods for the Type I MPA and Type I WMPA using the Akaike information criterion (AIC) \cite{akaike1974new} and Bayesian information criterion (BIC) \cite{schwarz1978estimating}.

To obtain the MLEs for both the Type I WPA and Type I WMPA, we set a tolerance level of $\epsilon=0.01$ between the $t^{th}$ and $\left(t+1\right)^{th}$ iterative values for each of the parameters. This tolerance level is used to determine the acceptable difference between the iterative values for each parameter. For the MoM estimates for the Type I WMPA, we have set the tolerance level as $\epsilon=0.001$ between the $t^{th}$ and $\left(t+1\right)^{th}$ iterative values for each of the parameters. We have chosen our values for $\epsilon$ with the same reasoning as in Section 6.

For this dataset, we can observe that the MLE method performs better than the MoM estimation method for both models.

\begin{table}[H]
		\caption{Comparison of estimation methods by AIC and BIC}
		\setlength{\tabcolsep}{8pt}
		\begin{tabular}{lccccccc}
			\noalign{\smallskip}\hline\noalign{\smallskip}  
			Model & Estimation Method & $\lambda_{1}$ & $\lambda_{2}$ & $\lambda_{3}$ & $\rho$ & AIC & BIC\\
			\noalign{\smallskip}\hline\noalign{\smallskip} 
			Type I MPA & MoM & 0.0544 & 0.3978 & 0.1303 & 0.3879 & 19984.48 & 20010.70 \\
			\textbf{Type I MPA} & \textbf{MLE} & \textbf{0.0930} & \textbf{0.4148} & \textbf{0.1257} & \textbf{0.3479} & \textbf{19890.39} & \textbf{19916.61} \\
			Type I WMPA & MoM & 0.1076 & 0.7835 & 0.2731 & 0.3454 & 19965.39 & 19991.61 \\
   		\textbf{Type I WMPA} & \textbf{MLE} & \textbf{0.1832} & \textbf{0.7714} & \textbf{0.2382} & \textbf{0.3328} & \textbf{19883.33} & \textbf{19909.55}\\
			\noalign{\smallskip}\hline\noalign{\smallskip} 
		\end{tabular}
\end{table}

In Table 5, we compare the fit of the Type I MPA and Type I WMPA using the MLEs for the parameters with the BP, DIBP, Type I MZIP, and the MZOIP using the values of  AIC and BIC as obtained in \cite{zhang2020multivariate}.

Both the Type I MPA and Type I WMPA demonstrate superior suitability for this dataset compared to the other models examined, with the Type I WMPA exhibiting the most optimal fit.

Upon calculating the $\widehat{\text{GDI}}$ for this dataset, we observe that the $\widehat{\text{GDI}}$ value is 2.72, which exceeds the threshold of 1. This indicates that the dataset is significantly overdispersed. It is therefore unsurprising that the Type I WMPA provided the best fit, given that it was constructed to account for more significant overdispersion.

\begin{table}[H]
		\caption{Comparison of AIC and BIC across 6 models}
		\setlength{\tabcolsep}{48pt}
		\begin{tabular}{lcc}
			\noalign{\smallskip}\hline\noalign{\smallskip}  
			Model & AIC & BIC\\
			\noalign{\smallskip}\hline\noalign{\smallskip} 
			BP & 22542.71 & 22562.38 \\
			DIBP & 20529.92 & 20556.14 \\
			Type I MZIP & 20565.82 & 20585.48  \\
   	    MZOIP & 20173.56 & 20212.89 \\ 
            \noalign{\smallskip}\hline\noalign{\smallskip}
   		\textbf{Type I MPA }& \textbf{19890.39} &          
            \textbf{19916.61} \\
            \textbf{Type I WMPA} & \textbf{19883.33} & \textbf{19909.55} \\
			\noalign{\smallskip}\hline\noalign{\smallskip} 
		\end{tabular}
\end{table}

\subsection{Example 2}

In Table 6 we consider a zero-inflated bivariate dataset based on an automobile portfolio from a major insurance company operating in Spain in 1995. The whole dataset consists of 80,994 policyholders. This dataset has been used extensively as an example of zero-inflated count data to measure the fit of multivariate zero-inflated models. A comparative analysis was performed in \cite{zhang2022comparative} to evaluate the performance of all the models that have been fitted to this dataset. Here, $N_1$ represents the number of third-party liability claims and $N_2$ represents all other automobile insurance claims.

\begin{table}[H]
		\caption{1995 automobile portfolio of a major insurance company operating in Spain}
		\setlength{\tabcolsep}{15pt}
		\begin{tabular}{lcccccccccc}
			\noalign{\smallskip}\hline\noalign{\smallskip}  
			$N_1/N_2$ & 0 & 1 & 2 & 3 & 4 & 5 & 6 & 7 & Total\\
			\noalign{\smallskip}\hline\noalign{\smallskip} 
			0 & 71087 & 3722 & 807 & 219 & 51 & 14 & 4 & 0 & 75904 \\
			1 & 3022 & 686 & 184 & 71 & 26 & 10 & 3 & 1 & 4003 \\
			2 & 574 & 138 & 55 & 15 & 8 & 4 & 1 & 1 & 796 \\
   		3 & 149 & 42 & 21 & 6 & 6 & 1 & 0 & 1 & 226 \\
   		4 & 29 & 15 & 3 & 2 & 1 & 1 & 0 & 0 & 51 \\
   		5 & 4 & 1 & 0 & 0 & 0 & 0 & 2 & 0 & 7 \\
   		6 & 2 & 1 & 0 & 1 & 0 & 0 & 0 & 0 & 4 \\
   		7 & 1 & 0 & 0 & 1 & 0 & 0 & 0 & 0 & 2  \\
   		8 & 0 & 0 & 1 & 0 & 0 & 0 & 0 & 0 & 1  \\
			Total & 74868 & 4605 & 1071 & 315 & 92 & 30 & 10 & 3 & 80994 \\
			\noalign{\smallskip}\hline\noalign{\smallskip} 
		\end{tabular}
\end{table}

In Table 7 we compare estimation methods for the Type I MPA and Type I WMPA using AIC and BIC. For the MLEs of both the Type I WPA and Type I WMPA, we set the tolerance level $\epsilon=0.01$ between the $t^{th}$ and $\left(t+1\right)^{th}$ iterative values for each of the parameters. As before, for the MoM estimates for the Type I WMPA we set the tolerance level $\epsilon=0.001$ between the $t^{th}$ and $\left(t+1\right)^{th}$ iterative values for each of the parameters. We have chosen our values for $\epsilon$ with the same reasoning as in Example 1 and Section 6.

As was the case in Example 1, for this dataset, we also observe that the MLE method performs better than the MoM estimation method for both models.

\begin{table}[H]
		\caption{Comparison of estimation methods by AIC and BIC}
		\setlength{\tabcolsep}{8pt}
		\begin{tabular}{lccccccc}
			\noalign{\smallskip}\hline\noalign{\smallskip}  
			Model & Estimation Method & $\lambda_{1}$ & $\lambda_{2}$ & $\lambda_{3}$ & $\rho$ & AIC & BIC\\
			\noalign{\smallskip}\hline\noalign{\smallskip} 
			Type I MPA & MoM & 0.0465 & 0.0631 & 0.0160 & 0.2274 & 96262.97 & 96300.18 \\
			\textbf{Type I MPA} & \textbf{MLE} & \textbf{0.0516} & \textbf{0.0658} & \textbf{0.0132} & \textbf{0.2150} & \textbf{96183.72} & \textbf{96220.93} \\
			Type I WMPA & MoM & 0.0981 & 0.1018 & 0.0256 & 0.3061 & 97029.90 & 97067.11 \\
   		\textbf{Type I WMPA} & \textbf{MLE} & \textbf{0.1022} & \textbf{0.1303} & \textbf{0.0265} & \textbf{0.2086} &      \textbf{96187.71} & \textbf{96224.92}\\
			\noalign{\smallskip}\hline\noalign{\smallskip} 
		\end{tabular}
\end{table}

In Table 8, using the MLEs for the parameters, we compare the performance of the Type I MPA and Type I WMPA with 10 other models excluding covariates in a comparative analysis as performed in \cite{zhang2022comparative}.

It is evident that both the Type I MPA and Type I WMPA exhibit superior performance to this dataset compared to the other models examined in \cite{zhang2022comparative}, with the Type I MPA performing the best. Calculating the $\widehat{\text{GDI}}$ for this dataset, we find that $\widehat{\text{GDI}}=1.88>1$. It is evident that this dataset exhibits overdispersion, although to a lesser extent than the dataset discussed in Example 1. Consequently, it is reasonable to expect that the Type I MPA would yield a more suitable fit in this scenario compared to the Type I WMPA.

\begin{table}[H]
		\caption{Comparison of AIC and BIC across 12 models excluding covariates}
		\setlength{\tabcolsep}{22pt}
		\begin{tabular}{lcccc}
			\noalign{\smallskip}\hline\noalign{\smallskip}  
			Model & Parameters & LogLik & AIC & BIC\\
			\noalign{\smallskip}\hline\noalign{\smallskip} 
			MIP & 2 & -53,271.05 & 106,546.10 & 106,564.70 \\
			MINB & 4 & -48,949.67 & 97,907.34 & 97,944.55 \\
			MIH & 6 & -48,948.02 & 97,908.03 & 97,963.85 \\
   		MP & 3 & -52,283.93 & 104,573.90 & 104,601.80 \\
   		MNB & 3 & -48,314.53 & 96,635.06 & 96,662.97 \\
            \noalign{\smallskip}\hline\noalign{\smallskip}
   		Type I MZIP & 3 & -48,630.52 & 97,267.03 & 97,294.94 \\
   		Type I MZINB & 5 & -48,101.02 & 96,212.03 & 96,258.54 \\
   		Type I MZIH & 7 & -48,087.96 & 96,189.91 & 96,255.03 \\
   		Type II MZIP & 4 & -48,630.52 & 97,269.03 & 97,306.24 \\
			Type II MZINB & 4 & -48,310.44 & 96,628.88 & 96,666.09 \\
            \noalign{\smallskip}\hline\noalign{\smallskip}
            \textbf{Type I MPA} & \textbf{4} & \textbf{-48087.86} & \textbf{96183.72} & \textbf{96220.93} \\
   		\textbf{Type I WMPA} & \textbf{4} & \textbf{-48089.86} &     \textbf{96187.71} & \textbf{96224.92} \\
			\noalign{\smallskip}\hline\noalign{\smallskip} 
		\end{tabular}
\end{table}

\section{Conclusion}\label{sec:8}

In this paper we introduced two new multivariate count data models, the Type I multivariate Pólya-Aeppli distribution as well as the Type I weighted multivariate Pólya-Aeppli distribution. We discussed properties of both distributions and we presented two different methods of parameter estimation for each. We also noted that both distributions are overdispersed and considered their performance on overdispersed zero-inflated and zero-and-one inflated count data. We observed that in both instances considered, these models provided a better fit to the Australian health survey and 1995 automobile insurance datasets than any of the existing models that it was compared to.

Some limitations with multivariate Pólya-Aeppli distributions is that they assume a positive correlation structure similar to the multivariate Poisson distribution, as is the case for all distributions constructed from Poisson. In addition, when modelling zero-inflated and zero-and-one inflated multivariate count data, the dispersion has to be sufficiently large in order for these models to provide a good fit. However this is often the case in practice.

In summary, we assert that these models make a substantial contribution to the existing body of
literature and the modeling of overdispersed, zero-inflated, and zero-and-one inflated multivariate count data.

\section{Appendix}

\subsection{Mathematical Results}

The following well known mathematical results are used throughout
this paper and referred to within the relevant sections.

\begin{flalign}
\begin{alignedat}{1}\sum_{k=0}^{\infty}z{}^{k} & =\frac{1}{1-z}\\
\sum_{k=0}^{\infty}\left(\begin{array}{c}
k+\beta\\
k
\end{array}\right)z^{k} & =\frac{1}{\left(1-z\right)^{\beta+1}}\\
\sum_{y=0}^{\infty}\frac{\left(\lambda s\right)^{y}}{y!} & =e^{\lambda s}\\
\sum_{y=0}^{\text{\ensuremath{\infty}}}\frac{e^{-\lambda}\lambda^{y}}{y!} & =1\\
\sum_{k=0}^{\infty}\sum_{i=0}^{\infty}a_{k.i} & =\sum_{i=0}^{\infty}\sum_{k=0}^{i}a_{k.i-k}\\
\sum_{k=0}^{\infty}\sum_{i,j=0}^{\infty}a_{k.i..j} & =\sum_{i,j=0}^{\infty}\sum_{k=0}^{min\left(i,j\right)}a_{k.i.-k.j-k}\\
\sum_{i=0}^{\infty}\sum_{k=i}^{\infty}a_{i.k} & =\sum_{k=0}^{\infty}\sum_{i=0}^{k}a_{i.k}\\
\sum_{i=0}^{\infty}\sum_{k=0}^{\infty}a_{i.k} & =\sum_{k=0}^{\infty}\sum_{i=0}^{\infty}a_{i.k}
\end{alignedat}
\label{A1}
\end{flalign}

\subsection{Laguerre Polynomials}

Laguerre and associated Laguerre polynomials are used throughout this
paper to simplify the calculation of the PMFs
of the various distributions and weighted distributions. The generating
function of the associated Laguerre polynomials is given by

{\large{}
\begin{equation}
\begin{alignedat}{1}\frac{1}{\left(1-z\right)^{\alpha+1}}e^{\left(\frac{xz}{z-1}\right)} & =\sum_{m=0}^{\infty}L_{m}^{\alpha}\left(x\right)z^{m}\end{alignedat}
\label{A26}
\end{equation}
}{\large\par}

with $\alpha\geq0$. The final form of the associated Laguerre polynomials given as

{\large{}
\begin{equation}
\begin{alignedat}{1}L_{n}^{\alpha}\left(x\right) & =\sum_{m=0}^{n}\left(-1\right)^{m}\left(\begin{array}{c}
n+\alpha\\
n-m
\end{array}\right)\frac{x^{m}}{m!}\end{alignedat}
.\label{A27}
\end{equation}
}{\large\par}

Following from (\ref{A27}), we can then easily calculate the below
results

{\large{}
\begin{equation}
\begin{alignedat}{1}L_{0}^{0}\left(x\right) & =1\\
L_{1}^{0}\left(x\right) & =1-x.
\end{alignedat}
\label{A28}
\end{equation}
}{\large\par}

From the following well known results

{\large{}
\begin{equation}
\begin{alignedat}{1}L_{n}^{\alpha}\left(x\right) & =\left(\frac{\alpha+1-x}{n}\right)L_{n-1}^{\alpha+1}\left(x\right)-\frac{x}{n}L_{n-2}^{\alpha+2}\left(x\right)\\
nL_{n}^{\alpha}\left(x\right) & =\left(n+\alpha\right)L_{n-1}^{\alpha}\left(x\right)-xL_{n-1}^{\alpha+1}\left(x\right)
\end{alignedat}
\label{eq:A29}
\end{equation}
}{\large\par}

we can see that

{\large{}
\begin{equation}
\begin{alignedat}{1}L_{n-1}^{\alpha}\left(x\right) & =\left(\frac{\alpha+1}{\alpha+n}\right)L_{n-1}^{\alpha+1}\left(x\right)-\frac{x}{\left(\alpha+n\right)}L_{n-2}^{\alpha+2}\left(x\right)\end{alignedat}
\label{eq:A30}
\end{equation}
}{\large\par}

and using \ref{eq:A29} and \ref{eq:A30}, it follows that

{\large{}
\begin{equation}
\begin{alignedat}{1}\frac{x^{2}}{\left(n+1\right)}L_{n-1}^{2}\left(x\right) & =\left(1-x\right)L_{n}^{0}\left(x\right)-L_{n+1}^{0}\left(x\right)\\
\frac{\left(-x\right)}{n}L_{n-1}^{1}\left(x\right) & =L_{n}^{0}\left(x\right)-L_{n-1}^{0}\left(x\right).
\end{alignedat}
\label{eq:A31}
\end{equation}
}{\large\par}

\section*{Acknowledgements}
This work was based upon research supported in part by the National Research Foundation (NRF) of South Africa (SA), grant ref SRUG2204203865 and RA171022270376 Nr 119109. The opinions expressed and conclusions arrived at are those of the authors and are not necessarily to be attributed to the NRF. 
\appendix

\bibliographystyle{elsarticle-num} 
\bibliography{database}


%% \begin{thebibliography}{00}



%% \end{thebibliography}
\end{document}